\documentclass[onecolumn,secnumarabic,amssymb,aps,praWA6M/00395330/6,superscriptaddress,longbibliography,a4paper,10pt]{revtex4-2}
\usepackage[utf8]{inputenc}
\usepackage{amsmath,amsthm,latexsym,amssymb,amsfonts,mathtools,epsfig}
\usepackage{graphicx,texdraw,subfigure,subeqnarray,fancyhdr,amsmath,multirow,color,bm,mathrsfs,xcolor}
\usepackage[english]{babel}
\usepackage[rightcaption]{sidecap}			
\usepackage{overpic}
\usepackage{microtype}
\usepackage{tikz}
\usepackage{lipsum}
\usepackage{url}            
\usepackage{upgreek}
\usepackage[colorlinks,citecolor=blue]{hyperref}
\newcommand{\comment}[1]{}

\newcommand{\pd}[2]{\frac{\partial #1}{\partial #2}} 
\newcommand{\td}[2]{\frac{\mathrm{d} #1}{\mathrm{d} #2}} 
\newcommand{\de}{\text{d}}

\begin{document}

\title{Twist and turn. Elastohydrodynamics of microscale active fibres}

\author{Maciej Lisicki}
\email{mklis@fuw.edu.pl}

\affiliation{Faculty of Physics, University of Warsaw, Pasteura 5, 02-093 Warsaw, Poland}

\begin{abstract}
    {Cellular locomotion often involves the motion of thin, elastic filaments, such as cilia and flagella, in viscous environments. The manuscript serves as a general introduction to the topic of modelling microscale elastohydrodynamics. We briefly characterise the specific features of biological filaments that affect their propulsion modes, and discuss the theoretical framework for their description, along with selected biological and artificial examples of active systems.} 
\end{abstract}

\maketitle

\section{Introduction}\label{sec1}

The purpose of this paper is to serve as a gentle introduction to the topic. The manuscript is based on a lecture given at the 2023 Geilo School {\it The Physics of Evolving Matter: Connectivity, Communication and Growth} under the same title.  On the one hand, we outline the biological complexity of the world of microswimmers and discuss the underlying molecular mechanisms that drive their active motion in fluids. On the other hand, we provide a unifying physical picture of Stokes flows, which pose universal limitations to the motion of microorganisms. As a result, we present a set of theoretical techniques that can be used to describe the motion of cilia and flagella in viscous environments. The framework can also be used to explore the dynamics of artificial microswimmers encompassing various driving mechanisms, provided that they involve slender structures. We present a selection of examples where such an approach has been useful for quantitative modelling.

Both hydrodynamics and elasticity can be unified as formal continuum theories relating stresses arising in a material in response to deformation (elasticity) or flow (hydrodynamics), and in this spirit they are discussed in numerous textbooks~\cite{truesdell1952mechanical,Lautrup2011,Landau1984-gr,Landau1987-pe}. In this paper, however, we shall turn to a less formal and more practical description of the popular tools used to describe the dynamics of elastic filaments in viscous fluids. We aim to present a spectrum of modelling strategies that we hope could inspire new questions or provide an overview of the field.

\section{Biological flagella and cilia}\label{sec:bio}

\begin{figure}[t]
    \centering
    \includegraphics[width=0.8\linewidth]{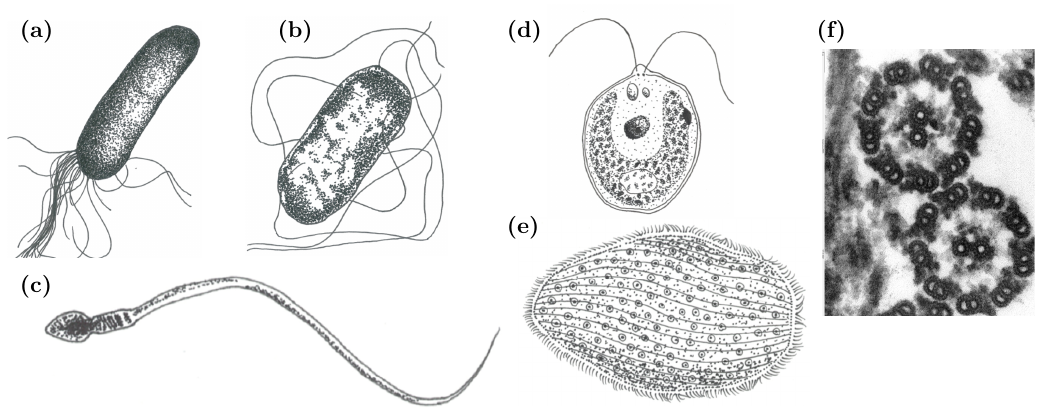}
    \caption{Selected examples of morphology of microswimmers.  {\bf (a)} Archaeon {\it Helicobacter pylori} has multiple passive flagella that form a bundle at the cap of the cell body. {\bf (b)} Bacterium {\it Escherichia coli} has several flagella distributed across the cell surface. {\bf (c)} Human spermatozoon has a single eukaryotic flagellum that performs wave-like motion. {\bf (d)} Biflagellate green alga {\it Chlamydomonas reinhardtii} beats its flagella in a breast-stroke-like manner. {\bf (e)} Ciliate {\it Opalina ranarum} has its cell body covered with thousands of cilia that exhibit metachronal waves. Sketches (not to scale) courtesy of Marcos F. Velho Rodrigues. {\bf (f)} Electron microscope photo of cross-sections of two eukaryotic flagella reveals the 9+2 microtubule scaffolding. Photo from OpenStax College, Biology (CC BY 3.0 licence).}\label{fig:bacteria}
\end{figure}

Life on Earth encompasses millions of species that exhibit an incredible diversity of form and function. Some of them are sketched in Fig.~\ref{fig:bacteria}. They can be divided into two domains: prokaryotes––single-cell organisms that do not have a nucleus, and eukaryotes whose cells possess nuclei. The domain of prokaryotes encompasses two kingdoms: bacteria and archaea, many of which are capable of locomotion and use their flagella to prey and navigate aquatic ecosystems. Eukaryotic organisms belong to one of four kingdoms: protists, animals, fungi, and plants. We shall focus on the microscale motility of protists, which contain simple protozoa and unicellular and multicellular algae that use their cilia for propulsion or pumping of the surrounding fluid. An important example of microscale motility in the animal kingdom are spermatozoa, whose activity is vital for reproduction. 

Both prokaryotes and eukaryotes use slender filamentous structures to survive in fluidic environments \cite{Sleigh,Brennen1977}. The fluid thus becomes the common denominator for nearly all microscale activity, and the physical laws of fluid dynamics pose universal limitations to the motion of microorganisms, which might involve swimming, stirring, or pumping the fluid. In all such problems, the small sizes and speeds of associated flows render fluid viscosity the dominant factor that determines the motion, as we discuss in the next Section.  

While the laws of physics provide certain universality to the domain of microscale locomotion, the diversity of shapes and motions of swimmers results in a plethora of interesting behaviours. At the root of them, the biological activity of slender filaments provides driving forces that induce fluid flow. However, the source of this activity is different in the two main domains of life.

\begin{figure}
    \centering
    \includegraphics[width=0.8\linewidth]{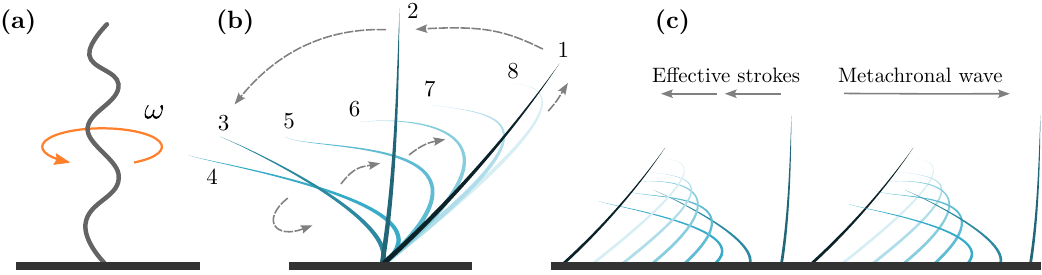}
    \caption{Beating patterns of active propulsion {\bf (a)} Most bacteria rotate their helical filaments by applying torque to the rotary motor that connects the cell body to the flagellum. The rotational motion of the helix induces translation of the cell. {\bf (b)} Active beating pattern of an individual cilium in the form of a two-stroke beat. Subsequent arrangements of the cilium are numbered and coloured, with the faintest colour at the latest time. The power stroke $1\to4$ is followed by a recovery stroke $4\to8$. {\bf (c)} Concerted motion of many cilia takes the form of a metachronal wave. One of its forms –– antiplectic –– travels in the direction opposite to the effective stroke direction. Redrawn after \cite{Brennen1977,Lauga2020}.}
    \label{fig:beating}
\end{figure}

Within prokaryotes, we shall restrict our attention to bacteria, which are much better understood and described than archaea. Bacteria swim by rotating their helical appendages~\cite{BERG1973}. They possess from one to a handful of flagella which they rotate in fluid. Each flagellum comprises a flagellar filament––an inert, helical biopolymer made of flagellin, typically longer than the cell itself (several micrometres), with a thickness of about 20 nm. The filament is connected to the cell body by an elastic joint, called the flagellar hook. The hook is then attached to the bacterial rotary motor, a structure embedded in the cell wall and driven by ion fluxes, which provides rotation of the filaments by application of torque. The typical applied torques are in the range of $10^3\ \text{pN}\,\text{nm}$, and result in rotational frequencies up to 300 Hz. The composite system of motor, hook, and flagellar filament is jointly referred to as the bacterial flagellum. Although some bacteria are powered by a single flagellum, the majority possess multiple flagella, each operated separately, distributed on the cell bodies. The helical shapes of flagella can take one of 11 polymorphic forms that differ in pitch angle and radius and depend on environmental conditions around the cell. The 'normal' polymorphic form has a radius of about $200\ \text{nm}$ and a pitch (wavelength measured around the helix axis) of about $2\ \mu\text{m}$. Except for rotation induced by the motors, bacterial flagella are passive and rotate in fluid almost like rigid bodies, as sketched in Fig.~\ref{fig:beating}a. A typical example of a prokaryotic microswimmer is the bacterium {\it E. coli}, which exhibits the so-called run-and-tumble motility \cite{Berg2004}. In the run phase, when the cell swims along straight trajectories, all flagellar motors rotate in the same direction, and the flagella bundle together. When one of the motors switches its direction of rotation, the resulting unbundling leads to a random reorientation of the cell, termed the tumble. After that, all motors rotate in the same direction again and the next run begins. 

In contrast, eukaryotic cells have active, flexible flagella which they actuate to generate stresses on the surrounding fluid. Their flagella are much thicker than their prokaryotic counterparts and have an intricate internal arrangement, termed the 9+2 structure. A cross section of an eukaryotic flagellum in Fig.~\ref{fig:bacteria}f reveals 9 microtubule doublets placed around the circumference and surrounding the central pair. This may be thought of as a scaffolding within the filament. Motor proteins (dyneins) cling to and walk along the microtubules, making them slide past each other. The resulting active stress distribution along the flagellum leads to a wide spectrum of motion seen in eukaryotic cells, including travelling waves, back-and-forth beating, planar and three-dimensional strokes \cite{Jahn1972}. This has been explored in depth in monoflagellate spermatozoa. Other well-studied examples of flagellated eukaryotes are biflagellate green algae of the genus {\it Chlamydomonas} \cite{Goldstein2015}. Individual cells can show a multitude of beating patterns but perhaps the most popular is the two-stroke motion. A cilium first performs a power stroke, moving in the plane normal to the cell body and pushing the fluid. Then, a recovery stroke follows, when the cilium moves closer to the cell wall and returns to its initial position, as illustrated in Fig.~\ref{fig:beating}b. Algal cells are also known to form multicellular and thus multiciliated colonies, such as those of {\it Volvox carteri} \cite{Brumley2012}. The presence of dense layers of eukaryotic flagella is characteristic for ciliates, a phylum encompassing ca. 4500 species of eukaryotes, such as {\it Paramecium}. Ciliates possess hundreds to thousands of relatively short flagella, called cilia, that perform their asymmetric beating in a concerted way. Their coordinated motion along the cell body, resembling a stadium wave, takes the form of metachronal waves. Metachronal waves can either be symplectic, when the wave travels in the direction of the power stroke of the cilia, or antiplectic, when these directions are opposite, as observed for {\it Paramecium} and depicted in Fig.~\ref{fig:beating}c.  This actuation can lead to complex helical motion of their host cell.  In result, such organisms create surface flows which drive the motion of surrounding fluid by the activity of their boundaries, represented using the famous squirmer model~\cite{Ishikawa2024}.

\section{Low Reynolds number flows}\label{sec:stokes}

\subsection{Stokesian hydrodynamics} In a continuum description of fluid flow, the central quantity of interest is the velocity field $\bm{v}\equiv\bm{v}(\bm{r},t)$. Its evolution is governed by the celebrated Navier-Stokes equations, which for an incompressible, Newtonian fluid of constant density $\rho$ and kinematic viscosity $\nu$ in the absence of body forces take the form
\begin{align}
\pd{\bm{v}}{t} + (\bm{v}\cdot\bm{\nabla})\bm{v} &= -\frac{1}{\rho} \nabla p + {\nu}\nabla^2 \bm{v},\label{eq:NS}\\
\nabla\cdot\bm{v} &= 0,
\end{align}
where $p$ denotes the pressure field. 
The latter equation expresses incompressibility and is derived from the mass conservation principle, while the former expresses momentum balance for a fluid element. Kinematic viscosity has units of stokes  (St, named after G.G. Stokes), with $1\ \text{St} = 10^{-4}\ \text{m}^2/\text{s}$ in SI units. Kinematic viscosity is related to shear, or dynamic, viscosity $\mu = \nu \rho$. The unit of $\mu$ is poise (P, named after L.M. Poiseuille), with $1\ \text{P} = 0.1\ \text{Pa}\cdot\text{s}$ in SI. For convenience, we note that the viscosity of water at 20$^\circ\text{C}$ is about 1 cP (dynamic), or 1 cSt (kinematic).

The full Navier-Stokes equations incorporate the effects of fluid inertia, embodied in the nonlinear term, and viscous stresses, represented by the Laplacian term. We can assess the mutual importance of these terms by considering a particular physical situation. We shall focus on flows that occur over a typical length scale of $L$, and with typical velocities of the order of $U$. The relative magnitude of the inertial and viscous terms is then expressed by the dimensionless Reynolds number
\begin{equation}
    \text{Re} = \frac{U L}{\nu} \sim \frac{|\bm{v}\cdot\bm{\nabla}\bm{v}|}{|\nu \nabla^2 \bm{v}|}. 
\end{equation}
The typical sizes of microswimmers ($L\sim 1$ m) and associated swimming speeds in aquatic environments ($U\sim 1$  $\upmu$m/s) yield the Reynolds number of the order of $10^{-3}$ or smaller. To this realm of low Reynolds number flows, also called Stokes flows, belong nearly all microscale flows, such as in microfluidic chips, colloidal suspensions, and slow, viscous flows. This points to an important intuition: To understand microscale flows, one can think of human-scale flows of high viscosity fluids, such as honey or syrup.

For vanishingly small Reynolds number, one can neglect the nonlinear term in Eq.~\eqref{eq:NS}. In addition, in the absence of intrinsic ultra-short time scales, the time derivative term will also be negligible, and the description reduces to the Stokes equations, or creeping flow equations
\begin{align}
- \nabla p + {\mu}\nabla^2 \bm{v} &= 0,\label{eq:St}\\
\nabla\cdot\bm{v} &= 0,
\end{align}
which are linear and stationary. This fact bears important consequences for microscale flows. Firstly, they are instantaneous, meaning that the flow depends only on the instantaneous configuration of the system (velocity, pressure, and therefore also stress in the fluid), and not on its history. Note that the motion of suspended particles in Stokes flow can be time-dependent, with the velocity field changing in time. At any instant, however, the net force and torque on each particle and each fluid element are zero. On the time scales of swimming, flow and pressure fields adjust instantaneously to the moving boundaries and particle surfaces. It follows from the Stokes equations \eqref{eq:St} that if a particle is moving under the action of external force $\bm{F}$ and torque $\bm{T}$, these are balanced by the corresponding hydrodynamic drag force $\bm{F}_\text{h}$ and torque $\bm{T}_\text{h}$, according to
\begin{align}
 \bm{F}_\text{h}+\bm{F} &= 0,\label{eq:F}\\
 \bm{T}_\text{h}+\bm{T} &= 0.\label{eq:T}
\end{align}
Secondly, it follows from the linearity of Stokes equation that the resulting velocity $\bm{U}$ and angular velocity $\bm{\Omega}$ of the particle are linearly related to the driving force and torque through the mobility matrix $\bm{\mu}$ \cite{KimKarrila}
\begin{equation}\label{eq:mobilitymatrix}
    \begin{pmatrix}
        \bm{V} \\
        \bm{\Omega}
    \end{pmatrix}
    = 
  \bm{\mu}
    \cdot
    \begin{pmatrix}
        \bm{F} \\
        \bm{T}
    \end{pmatrix} =  \begin{pmatrix}
        \bm{\mu}^{tt} & \bm{\mu}^{tr} \\
        \bm{\mu}^{rt} & \bm{\mu}^{rr}
    \end{pmatrix}
    \begin{pmatrix}
        \bm{F} \\
        \bm{T}
    \end{pmatrix} 
\end{equation}
where $\bm{\mu}$ is further split into translational ($tt$), rotational ($rr$) and translation-rotation coupling parts ($tr$ and $rt$). This notion can also be generalised to multiple immersed particles, and the description follows in terms of higher-dimensional matrix relations. The mobility matrix fully describes the effect of hydrodynamic interactions for a given configuration of the system. For the very simple case of motion of a single spherical particle of radius $a$ through a viscous fluid under the action of a force $\bm{F}$, the velocity is found from the Stokes law as $\bm{U} = \mu_0 \bm{F},$
with $\mu_0 = (6\pi \mu a)^{-1}$, which can also be formulated in terms of the familiar expression for the Stokes drag force 
\begin{equation}\label{eq:stokeslaw}
    \bm{F}_\text{h} = - 6\pi \mu a \bm{U},
\end{equation}
in agreement with Eq.~\eqref{eq:F}. The single-sphere mobility coefficient $\mu_0$ can be inverted to define the friction coefficient $\zeta_0 = \mu_0^{-1} = 6\pi \mu a$. The friction matrix $\bm{\zeta}$ is then defined by analogy as the inverse of the mobility matrix
\begin{equation}\label{eq:frictionmatrix}
    \bm{\zeta} = \begin{pmatrix}
        \bm{\zeta}^{tt} & \bm{\zeta}^{tr} \\
        \bm{\zeta}^{rt} & \bm{\zeta}^{rr}
    \end{pmatrix} = \begin{pmatrix}
        \bm{\mu}^{tt} & \bm{\mu}^{tr} \\
        \bm{\mu}^{rt} & \bm{\mu}^{rr}
    \end{pmatrix}^{-1} = \bm{\mu}^{-1}.
\end{equation}
There matrices are both symmetric and positive definite, which follows from the second law of thermodynamics \cite{KimKarrila}.

\subsection{Green's functions and flow singularities} 
\begin{figure}
    \centering
    \includegraphics[width=0.8\linewidth]{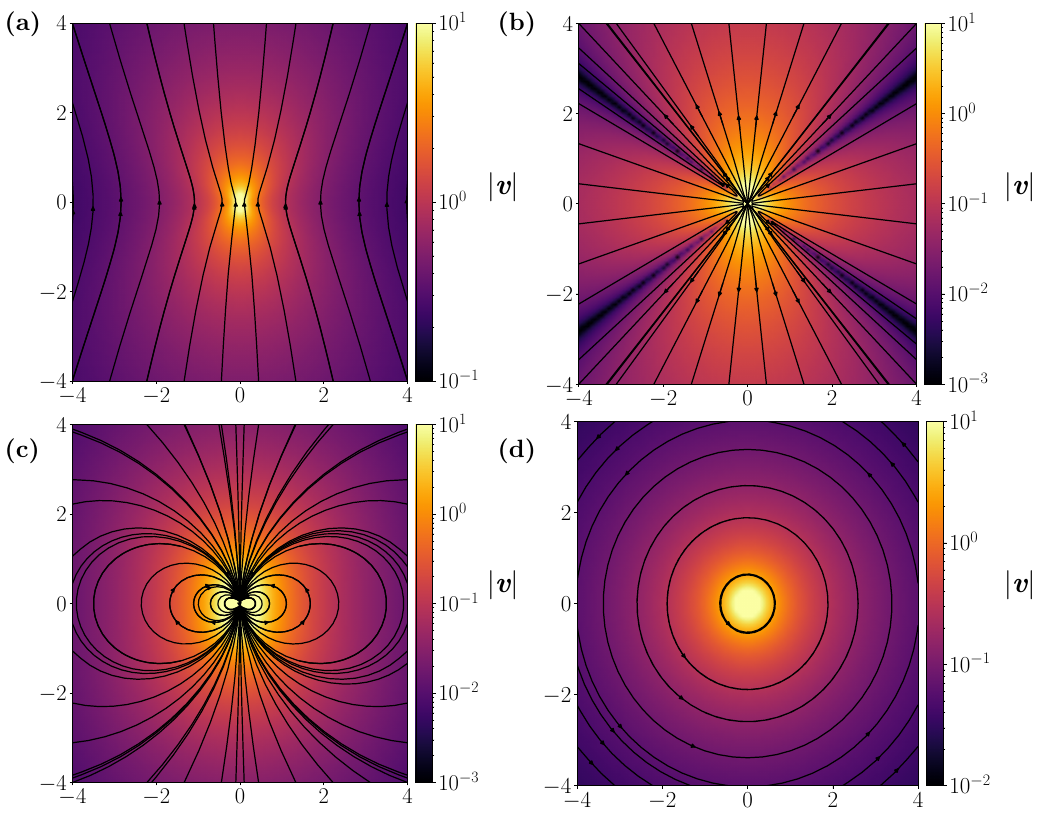}
    \caption{Streamlines of selected flow singularities with associated velocity magnitude marked in colour in logarithmic scale. {\bf (a)} Stokeslet flow, Eq.~\eqref{eq:stokeslet}, with the force pointing upwards, $\bm{e}=\bm{e}_y$. {\bf (b)} Stokes doublet (stresslet), Eq.~\eqref{eq:stresslet}, with $\bm{d}=\bm{e}=\bm{e}_y$. {\bf (c)} Source doublet, Eq.~\eqref{eq:doublet} with $\bm{e}=\bm{e}_y$. {\bf (d)} Rotlet field, Eq.~\eqref{eq:rotlet}, with $\bm{d}\times\bm{e}=\bm{e}_z$.}
    \label{fig:flows}
\end{figure}

The linearity of Stokes equations~\eqref{eq:St} offers the possibility to construct solutions by superposition of fundamental point-force flows. The Green's function of Stokes equations, known as the \emph{Stokeslet}, provides the flow caused by a concentrated force $\bm{F}$ acting at a point $\bm{r}_0$, and is given by
\begin{equation} \label{eq:stokeslet}
    \bm{v}(\bm{r}) = \bm{\mathsf{T}}(\bm{r}-\bm{r}_0)\cdot\bm{F},
\end{equation}
where the Oseen tensor $\bm{\mathsf{T}}(\bm{r})$ is given in terms of the distance from the point force as
\begin{equation}\label{eq:stokeslet2}
    \bm{\mathsf{T}}(\bm{r}) = \frac{1}{8\pi \mu r} (\bm{1} + \hat{\bm{r}}\hat{\bm{r}}),
\end{equation}
 where $r=|\bm{r}|$ and $\hat{\bm{r}} = \bm{r}/r$. Its Cartesian components are given by $\mathsf{T}_{ij}(\bm{r}) = (8\pi \mu r)^{-1} \left({\delta_{ij}} + {\hat{r}_i \hat{r}_j}\right)$, and the flow is sketched in Fig.~\ref{fig:flows}a. We note that the flow field due to a point force is anisotropic, with the velocity along the force direction being twice the value of the velocity in the perpendicular direction at a given distance. In the far field, the Stokeslet decays as $1/r$, with a notably long range.

 Another important observation is the fact that the derivatives of the Stokeslet also satisfy the Stokes equations~\eqref{eq:St}, and therefore are solutions that satisfy the same boundary conditions. An example of a solution which is relevant for the description of microswimmers is the {\it Stokes dipole}. It may be constructed in the following manner: Consider a pair of opposite point forces, $F\bm{e}$ located at the origin and $-F\bm{e}$, at a distance $\ell\bm{d}$ apart, with $\bm{d},\ \bm{e}$ being unit vectors. In the limit of $\ell\to 0$, the total dipole flow field $\bm{v}^d$ can then be approximated by
 \begin{equation}
     \bm{v}_\text{D}(\bm{r}) =  F \,\bm{\mathsf{T}}(\bm{r})\cdot \bm{e} - F\,\bm{\mathsf{T}}(\bm{r}-\ell\bm{d})\cdot\bm{e} \approx - \mathcal{P}(\bm{d}\cdot\nabla)\bm{\mathsf{T}}\cdot\bm{e},
 \end{equation}
 with the Stokes dipole moment $\mathcal{P} = \ell F$.
 The character of this velocity field depends on the relative orientation of the unit vectors $\bm{d}$, indicating the direction of separation of the points where the forces are applied, and $\bm{e}$ representing the direction along which the forces act. In order to interpret this expression, let us now take an alternative, more abstract look at this relationship. The first term in the Taylor expansion of the velocity field at any given point involves the gradient of the Stokeslet, $\nabla\bm{\mathsf{T}}$, which is a third-order tensor with three indices, $\partial_i \mathsf{T}_{jk}$, and with $\partial_i = \partial/\partial r_i$ denoting a spatial derivative. The velocity field may then be constructed by contracting this tensor with two vectors. This operation can be looked upon as a contraction with a second-order tensor $\bm{\mathsf{Q}}$ being a dyadic product of vectors $\bm{d}$ and $\bm{e}$, so that  $\bm{\mathsf{Q}}=\bm{d}\bm{e}$, or $\mathsf{Q}_{ij}= d_i e_j$. Using the Einstein summation convention, the general velocity field can then be written as
 \begin{equation}
     (v_\text{D})_j = \mathcal{P} d_i e_k \partial_i \mathsf{T}_{jk} = \mathcal{P} \mathsf{Q}_{ik} \partial_i \mathsf{T}_{jk}.
 \end{equation}
To gain further insight, we can decompose $\bm{\mathsf{Q}}$ into its trace $Q$, the traceless symmetric part $\bm{\mathsf{S}}$, and the antisymmetric part $\bm{\mathsf{A}}$ via
\begin{equation}
   Q = \text{Tr}(\bm{\mathsf{Q}}), \qquad \bm{\mathsf{S}} = \frac{1}{2}\left( \bm{\mathsf{Q}} + \bm{\mathsf{Q}}^T\right)-\frac{1}{3} Q\bm{1}, \qquad \bm{\mathsf{A}} = \frac{1}{2}\left( \bm{\mathsf{Q}} - \bm{\mathsf{Q}}^T\right),
\end{equation}
with $T$ denoting the transpose. The velocity field associated with the trace term vanishes because the divergence of the Oseen tensor vanishes, $\partial_j \mathsf{T}_{jk} = 0$. The symmetric part is called the \emph{stresslet}, and its velocity field can be written as
\begin{equation} \label{eq:stresslet1}
    \bm{v}_D^\text{S} = \frac{3\mathcal{P}}{8\pi\mu}\frac{\bm{r}\cdot{\bm{\mathsf{S}}}\cdot\bm{r}}{r^5},
\end{equation}
where its $i$-th component may be explicitly written using the arbitrary vectors $\bm{d}$ and $\bm{e}$ as
\begin{equation}\label{eq:stresslet}
    (\bm{v}_D^\text{S})_i = \frac{\mathcal{P}}{3\pi\mu} \left[-\frac{d_j e_j r_i}{r^3} + 3 \frac{e_j r_j d_k r_k r_i}{r^5}\right].
\end{equation}
Its velocity field for the particular case of $\bm{d}=\bm{e}$, also discussed later on, is depicted in Fig.~\ref{fig:flows}b. The antisymmetric part is called a \emph{rotlet}, with the corresponding flow field being
\begin{equation} \label{eq:rotlet}
    \bm{v}_D^\text{A} = \frac{\mathcal{P}}{8\pi\mu}\frac{(\bm{d}\times\bm{e})\times\bm{r}}{r^3},
\end{equation}
as drawn in Fig.~\ref{fig:flows}d. It can be interpreted as a flow created by a point torque applied to the fluid \cite{Lauga2020}.

One can extend this procedure and calculate higher-order gradients of the Stokeslet to find a representation of the flow field. One useful singularity is the potential source dipole, which is the flow field resulting from a sink and a source brought together. A source in Stokes flow placed at the origin has the flow field $\bm{v}_S(\bm{r})=-\bm{r}/4\pi \mu r^3$. A sink has the opposite direction of the flow field. Thus a source dipole field is given in terms of both force singularities and source singularities as
\begin{equation}\label{eq:doublet}
    \bm{v}_{SD} = \bm{e}\cdot\nabla \bm{v}_S(\bm{r}) =  \frac{1}{2}  \nabla^2 \bm{\mathsf{T}}(\bm{r})\cdot\bm{e} = \frac{1}{4\pi \mu}\frac{\bm{1} - 3\hat{\bm{r}}\hat{\bm{r}}}{r^3}\cdot{\bm{e}},
\end{equation}
 with the associated flow depicted in Fig.~\ref{fig:flows}c. For a detailed discussion of hydrodynamic singularities, we refer to the work of Chwang \& Wu \cite{Chwang1975}, and the discussion of its applications in the context of microswimmers \cite{Lauga2020,Spagnolie2012}.

 Within this framework, it is also possible to incorporate the effect of confinement for select geometries using a variant of the method of images. Exact solutions are known for highly symmetric geometries, such as that of a planar surface, inside or outside a rigid sphere or a cylindrical tube. We note here that the image system for a Stokeslet near a planar free interface is simply a mirror image Stokeslet. This guarantees that the velocity field normal to the surface vanishes and satisfies the no normal stress boundary condition. For a no-slip wall, or a fluid-fluid interface the image system of singularities is more complex, as derived by Blake \cite{Blake1971,aderogba1978action} for a Stokeslet parallel and perpendicular to the surface; For a Stokeslet at a distance $h$ from a rigid wall, the image system contains a Stokeslet, but also a Stokes dipole, and a source dipole, with their strengths dependent on $h$ and $h^2$, respectively, and orientations dependent on that of the original Stokeslet. Solutions have also been derived for the case when the interface is partially covered by surfactant~\cite{blawzdziewicz1999stokes}. In some cases, explicit expressions for higher order singularities have been derived, e.g. for a Stokes dipole outside a spherical shell \cite{Chamolly2020stokes}.

\subsection{Swimming in microscale}\label{sec:swimming}

Since the Reynolds number is small, Stokes equations~\eqref{sec:stokes} apply to the description of motion of swimming microorganisms~\cite{Lauga2020,Purcell1977,lighthill1976flagellar,Lauga:2009,Lauga2016,Pedley1992}. 
Their properties determine the rules which cellular microswimmers and their flagellar appendages have to obey. In the classical introduction to how life looks like in the microworld, Purcell discusses at length the consequences of Stokes flows for locomotion \cite{Purcell1977}. One of them is the kinematic reversibility of Stokes flow, which can be demonstrated in a macroscale experiment using a very viscous fluid to mimic the conditions of vanishingly small $\text{Re}$, as illustrated by G.I. Taylor in the televised lecture~\cite{Taylor,Taylor2}. The famous Scallop Theorem, saying that no time-reversible swimming gait can lead to net propulsion, outlines basic design rules for artificial microswimmers to follow. We note, however, that elasticity breaks this time-reversibility constraint \cite{Lauga2011}.

Stokes equations are instantaneous, reflecting the fact that the timescales of acceleration or deceleration of microswimmers are negligibly small compared to the timescales of motion, so that Newton's second law for any microswimmer can be formulated by neglecting the acceleration part. In consequence, the total force and torque acting on a swimming cell must vanish, as in Eqs.~(\ref{eq:F}) and~(\ref{eq:T}). For this reason, most swimming cells are both force- and torque-free. This means that their motion applies no net force or torque on the fluid and, therefore, their flow field cannot be represented by a Stokeslet~\eqref{eq:stokeslet} or a rotlet~\eqref{eq:rotlet}. If we stick to the simple picture presented above, the flow field far away from a swimmer should be accurately represented by a pair of collinear forces. This corresponds to a particular case of Eq.~\eqref{eq:stresslet1} when $\bm{d}=\bm{e}$, the stresslet tensor simplifies to $\bm{\mathsf{S}}=(\bm{e}\bm{e}-\tfrac{1}{3}\bm{1})$ and the resulting flow field becomes
\begin{equation} \label{eq{stresslet}}
    \bm{v}_D^\text{S} =\frac{\mathcal{P}}{3\pi\mu} \left[-\frac{1}{r^3} + 3 \frac{(\bm{e}\cdot\bm{r})^2}{r^5}\right]\bm{r}.
\end{equation}
Thus the most general flow field of a microswimmer will be that of a stresslet, which is purely radial with a direction-dependent magnitude and an asymptotic decay of $r^{-2}$. The dipole moment of a cell, $\mathcal{P}$, may in general depend on time. The sign of the dipole moment, $\mathcal{P}$, has important biophysical consequences. Cells with  $\mathcal{P}>0$ are called \emph{pushers}, and this category encompasses most bacteria and sperm cells. They use their flagella to 'push' the surrounding fluid along their swimming direction and drag it in from the sides. The second category are \emph{pullers}, which includes some algae, that 'pull' on the fluid at the front and the back, and stream it in the direction normal to their motion \cite{Lauga:2009}. Importantly, the flow fields of all swimmers are front-back symmetric, so the dipole moment is not related to the direction of swimming.

The far field dipole signature of flows generated by microswimmers has been corroborated in particle image velocimetry (PIV) experiments in prokaryotic bacterium {\it E. coli}~\cite{Drescher2011} and eukaryotic alga {\it Chlamydomonas}~\cite{Klindt2015}. The dipolar flow field decays with the distance $r$ from the swimmer as $r^{-2}$. Swimmers that do not produce such strong flows, i.e. their flow fields decay faster with distance, are referred to as {\it neutral} swimmers. Getting closer to the swimmer, one sees flow signatures that reflect the shape and swimming gait of the cell. The total flow can be represented by a combination of higher-order singularities, such as rotlet dipoles, Stokes quadrupoles, source quadrupoles, etc. \cite{Lauga2020}.

All swimming microorganisms achieve propulsion by changing the shapes of their bodies or their appendages. As a result, they achieve a time- and configuration-dependent linear and angular velocity, $\bm{U}(t)$ and $\bm{\Omega}(t)$. To find the velocity, one needs to specify the kinematics of a given swimmer, and use them as boundary conditions to Stokes equations~\eqref{eq:St} to find the stresses which the organism exerts on the fluid, which can then be integrated over the body surface to find the total thrust force $\bm{F}_t$ and torque $\bm{T}_t$. The thrust forces and torques are exactly balanced by the hydrodynamic drag and torque by Eqs.~(\ref{eq:F}) and~(\ref{eq:T}). Once the (configuration-dependent) mobility matrix for the organism is known, the swimming speed and angular velocity are found from Eq.~\eqref{eq:mobilitymatrix} as
\begin{equation}\label{eq:thrust}
    \begin{pmatrix}
        \bm{U}(t) \\
        \bm{\Omega}(t)
    \end{pmatrix}
    = 
  \bm{\mu}(t)
    \cdot
    \begin{pmatrix}
        \bm{F}_t(t) \\
        \bm{T}_t(t)
    \end{pmatrix}.
\end{equation}
A detailed discussion of this relationship can be found in Ref.~\cite{Lauga2020}.

\subsection{Hydrodynamics of thin filaments}

Elongated fibres are an important element of many fluid systems, relevant to biology, engineering, and physics applications. The presence of polymers in a Newtonian solvent endows it with non-linear rheological characteristics; swimming microorganisms use their slender appendages for propulsion; filamentous microtubules provide structural stability to cells and mediate active transport inside them \cite{Shelley2016rev}; DNA strands form supercoiled structures owing to their internal twist that changes their hydrodynamic properties in solution \cite{Waszkiewicz2023}; In these examples, and in many more, natural and synthetic flexible filaments of high slenderness move in a viscous environment, and fluid drag forces play and important role in shaping their dynamics.

Since resolving the full dynamics of a slender fibre would require a complex coupling between its internal elastic stresses and fluid stresses outside, numerous approximate techniques have been developed to aid modelling. 

\subsubsection{Bead models}

Bead models are the basis for many hydrodynamic applications. Before proceeding to the case of filaments, consider an arbitrary collection of identical spheres located at the points $\bm{r}_i$, with $i=1,\ldots,N$, and assume that the forces acting on the particles $\bm{F}_i$ are given. They might come from an external field, or from interactions between them. An isolated particle under the action of a force $\bm{F}$ would move with a velocity $\bm{U} = \mu_0 \bm{F}$, dictated by the Stokes law, Eq.~\eqref{eq:stokeslaw}. The presence of other particles affects this translational velocity. The mobility matrix of Eq.~\eqref{eq:mobilitymatrix} now connects all forces and torques with all velocities, both translational and rotational, and thus becomes an $6N \times 6N$ matrix, as discussed in detail in Refs.~\cite{Ekiel,Brady1988,Lisicki2016nagele}. Even if we restrict our attention only to translational motion, the translational mobility matrix $\bm{\mu}^{tt}$ is a $3N\times 3N$ tensor that depends on the positions of all particles. The velocity of sphere $i$ will then be given by
\begin{equation} \label{eq:hydromotion}
    \bm{U}_i = \sum_{j} \bm{\mu}^{tt}_{ij}\cdot \bm{F}_j = \mu_0 \bm{F}_i + \sum_{j\neq i} \bm{\mu}^{tt}_{ij}\cdot \bm{F}_j,
\end{equation}
where $\bm{\mu}_{ij}$ is a sub-matrix of $\bm{\mu}^{tt}$ that connect the particles $i$ and $j$, and we singled out the self-term $i=j$ being the Stokes velocity.

A convenient analytical approximation is to retain only far-field effects in hydrodynamic interactions. The resulting mobility matrices in the Rotne-Prager-Yamakawa (RPY) approximation~\cite{rotne1969,yamakawa1970,Dhont} involve pair-wise interactions, with
\begin{align} 
\bm{\mu}_{ii}^{tt,RP}  &= \mu_0 \bm{1}, \\
\bm{\mu}_{ij}^{tt,RP}  &= \mu_0 \left[ \frac{3}{4}  \frac{a}{r_{ij}} \left( \bm{1} + \bm{\hat{r}}_{ij} \bm{\hat{r}}_{ij} \right) + \frac{1}{2}  \frac{a^3}{r_{ij}^3} \left( \bm{1} -3  \bm{\hat{r}}_{ij}\bm{\hat{r}}_{ij} \right) \right],\quad \text{for}\ j\neq i,
    \label{eq:rotne-prager-translation}
\end{align}
where $\bm{r}_{ij} = \bm{r}_i - \bm{r}_j$, $r_{ij}=|\bm{r}_{ij}|$ and $\bm{\hat{r}}_{ij} = \bm{r}_{ij} / r_{ij}$.
These expressions have also been generalised to possibly overlapping spheres of unequal sizes \cite{Zuk2014}.

We may now think of a rod-shaped filament of length $L$ as a collection of beads of radius $a$, collinear with a unit director $\bm{{t}}$. One can then use the mobility matrix formalism to find the mobility (and friction) coefficients for a straight rod \cite{Dhont}. The particular geometric arrangement renders hydrodynamic friction on this complex particle anisotropic, so that under the action of a force $\bm{F}$ a rod will move with the velocity 
\begin{equation}
    \bm{V} = [\mu_\parallel \bm{{t}}\bm{{t}} + \mu_\perp (\bm{1}- \bm{{t}}\bm{{t}})]\cdot \bm{F},
\end{equation}
with the anisotropic mobility and friction coefficients
\begin{equation}\label{frictioncoeff}
    \mu_\parallel = \frac{1}{\zeta_\parallel} \approx \frac{\ln(L/2a)}{2\pi \mu L},  \qquad \mu_\perp =\frac{1}{\zeta_\perp} \approx \frac{\ln(L/2a)}{4\pi \mu L}.
\end{equation}
 Note that for the long and thin rods considered here, the parallel mobility coefficient is twice the perpendicular mobility coefficient, reflecting the hydrodynamic anisotropy of the system. Similar exact expressions are available for slender, prolate spheroids \cite{KimKarrila}. We shall see how this result emerges from the continuum theory in the next subsection.

\subsubsection{Continuum models: slender body theory} 

Resolving the motion of a filament in flow amounts to solving the Stokes equations~\eqref{sec:stokes} with the velocity of the fluid matching that of the filament on its surface, and appropriate flow at infinity. It can be formulated in terms of boundary integral equations \cite{Pozrikidis2002} which  involve a considerable computational burden to solve. However, the slenderness of the filaments can be used to reduce these equations to filament centre lines. Consider a centre line, whose position $\bm{r}(s)$ is parametrised by the arc length $s\in[0,L]$. We discuss the geometry of such centre lines in detail in Sec. \ref{sec:elast_fil_geom}. The idea behind the so-called \emph{slender body theory} (SBT) is to represent the flow around a filament by placing fundamental singularities---Stokeslets [Eq.~\eqref{eq:stokeslet}] and Stokes doublets---[Eq.~\eqref{eq:doublet}] on the centreline, and then to use the technique of matched asymptotic expansions to derive the approximate equation for their distribution. The procedure is subtle and we refer to the more detailed discussion, e.g. in Refs. \cite{Cox1970,lighthill1976flagellar,Batchelor1970,Keller1976a,Johnson1980a,koens_lauga_2018}. 

For a filament with tapered ends, and given the force density per unit length, $\bm{f}$, that the filament exerts on the fluid, and in the presence of external background flow $\bm{U}_0(s,t)$, the velocity of the fibre, $\bm{U}(s,t)=\partial \bm{r}(s,t)/\partial t$, is given by  
\begin{equation}
\label{eq:generalSBT}
	8\pi\mu [\bm{U}(s,t) - \bm{U}_0(s,t)]= -\bm{\Lambda}[\bm{f}](s) - \bm{K}[\bm{f}](s). 
\end{equation}
We now drop the arguments for brevity. The first term on the right-hand side is a local term $\bm{\Lambda}$, given by
\begin{equation}
	\bm{\Lambda}[\bm{f}] = [(c+1)\bm{1} + (c-3)\bm{t}\bm{t}]\cdot\bm{f},
\end{equation}
with the slenderness parameter $c$ determined by the body cross-sectional shape and radius \cite{koens_lauga_2018}, and the tangent vector is determined by $\bm{t}=\partial\bm{r}/\partial s$. The second term is the global contribution that captures the filament self-interactions
\begin{equation}
	\bm{K}[\bm{f}] = \int_{0}^{L}\left( \frac{\bm{1} + \hat{\bm{R}}\hat{\bm{R}}}{|\bm{R}|}\cdot\bm{f}(s') - \frac{\bm{1} + \bm{t}\bm{t}}{|s-s'|}\cdot\bm{f}(s)\right)\de s',
\end{equation}
where $\bm{R}(s,s') = \bm{r}(s)-\bm{r}(s')$ and $\hat{\bm{R}} = \bm{R}/|\bm{R}|$. The problem of singularity of the last integrand can be treated in a practical way proposed by Tornberg \& Shelley \cite{Tornberg2004} who introduced a cutoff of long-wavelength modes by regularising the denominator in the integrand. This is related to the general method of regularised Stokeslets, developed by Cortez {\it et al.}~\cite{Cortez}, where the singular kernels of fundamental flow solutions are regularised by adding a small constant in the denominators to allow numerical treatment. 

An important simplification emerges when the non-local term is neglected. The resulting equations are known as resistive force theory (RFT) of \citet{Gray:1955}. In RFT, the local velocity of the filament is proportional to the local force density, $\bm{f}$, by
\begin{equation}
\bm{U}(s) = \left[\mu_\parallel \bm{t}\bm{t} + \mu_\perp\left(\bm{1} - \bm{t}\bm{t}\right)\right]\cdot\bm{f}(s),\label{eq:local_sbt}
\end{equation}
with the mobility coefficients $\mu_\parallel = 2\mu_\perp = c / 4\pi\mu L$, which matches the results of the bead model for a particular case of $c=2\ln(L/2a)$.
RFT is a popular modelling technique in biological fluid dynamics, and has proved useful in the analysis  of experimental observations of deforming flagella \citep{lauga_eloy_2013},  buckling instabilities \citep{Decanio2017,Cholakova2021}, swirling of flexible microtubules in the cytoskeleton \citep{Stein2021}, or synchronisation and bundling in bacterial flagella \cite{Tatulea2022,Man_2016}. A detailed comparison between the predictions of RFT and those of SBT was presented by \citet{JohnsonBrokaw1979} and later \citet{Rodenborn2013}.

\section{Elasticity of thin filaments}

Here, we focus on two major categories of models used to encompass the elasticity of thin, filamentous structures. From the computational point of view, bead models are typically straightforward to implement numerically, yet they have relatively larger aspect ratios. On the other hand, theoretical models are more straightforward to formulate for very slender continuum filaments and can be discretised for numerical calculations. We present these two choices briefly below.

\begin{figure}
    \centering
    \includegraphics[width=0.8\linewidth]{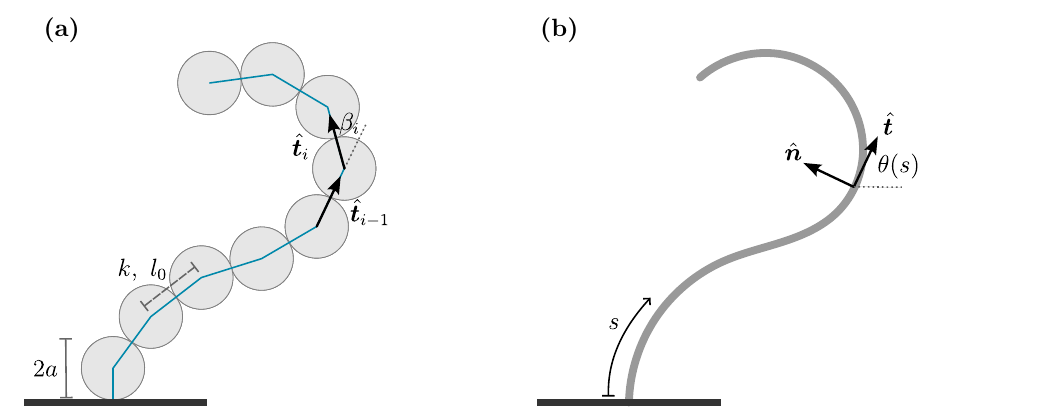}
    \caption{Geometry of two types of models of elastic thin filaments. {\bf (a)} Bead-model formed of spherical subunits connected with elastic springs of stiffness $k$ and equilibrium length $l_0$. Bending rigidity is implemented by imposing an energetic cost of having non-zero bending angles $\beta_i$ for each bead. {\bf (b)} Continuum model of a thin filament, parametrised by the arc length $s$, with spatially-dependent material frame composed (in 2D) of the tangent vector $\hat{\bm{t}}$ and normal vector $\hat{\bm{n}}$. The shape can also be parametrised by the tangent angle $\theta(s)$.}
    \label{fig:filaments}
\end{figure}

\subsection{Bead models}\label{sec:bead} 

An elastic fibre or polymer can be represented as a collection of spherical beads, resembling a string of pearls. By the addition of appropriate interaction potentials, we can endow such a fibre with the desired elastic properties, such as stretching and bending rigidity. For simplicity, we consider a chain of $N$ identical spherical beads of diameter $d=2a$ each. The centres of consecutive spheres are connected by springs, which have an equilibrium length of $l_0$, typically close to the bead diameter.

For an extensible chain, with the centres of bead located at $\bm{r}_i$, the stretching interaction potential $U_s$ is typically assumed to be harmonic, given by
\begin{equation}
    U_s = \frac{k}{2} \sum_{i=1}^{N-1}  \left(|\bm{r}_i-\bm{r}_{i+1}|-l_0\right)^2,
\end{equation}
with a constant spring stiffness $k$. Inextensibility may be simulated introducing a different interaction potential, e.g. FENE \cite{Shashank2023} or by introducing equal and opposite forces along the links between particles \cite{Lagomarsino2003}. The stretching force acting on bead $i$ is then calculated as
\begin{equation}
    \bm{F}^s_i = \pd{U_s}{\bm{r}_i}.
\end{equation}

Bending rigidity can be implemented using the notion of local bending angle $\beta_i$, being the angle between three consecutive beads, $i-1$, $i$, and $i+1$. Defining the vector connecting consecutive beads
\begin{equation}
    \bm{t}_i = \bm{r}_{i+1}-\bm{r}_{i},
\end{equation}
and an associated unit vector $\bm{\hat{t}}_i = \bm{t}_i / |\bm{t}_i|$, we can write the bending angle as
\begin{equation}
    \cos\beta_i = {\bm{\hat{t}}_{i-1}\cdot\bm{\hat{t}}_i}
\end{equation}
Historically, the idea of replacing a continuum slender filament by a discrete model dates back to Hencky \cite{Hencky,Wang2020}, who in 1920 proposed a bar-chain model composed of inextensible rigid segments of length $l_0$ each, connected by frictionless hinges with elastic torsional springs that provide the flexibility. Also called the Kratky-Porod model, it uses the following form for the bending potential energy
\begin{equation}
    U_b^{KP} = \frac{A}{2 l_0} \sum_{i=2}^{N-1} (\bm{\hat{t}}_i - \bm{\hat{t}}_{i-1})^2 =   \frac{A}{l_0} \sum_{i=2}^{N-1} (1 - \cos\beta_i),
\end{equation}
and has been mathematically proven to converge to Euler inextensible elastica in large deformation \cite{Alibert2017}.

A different stragihtforward possibility \cite{Saggiorato2015,Shashank2023} is the harmonic bending potential
\begin{equation}\label{eq:bendingharmonic}
    U_b^{H} = \frac{A}{2 l_0} \sum_{i=2}^{N-1} \beta_i^2.
\end{equation}
We assume that the bending stiffness $A$ and the spring constant $k$ are related to each other by the formula,
\begin{equation}
    k = \frac{4 A}{a^2 l_0},
\end{equation}
which comes from the analysis of an elastic beam of radius $a$ and Young's modulus $E_Y$, for which the bending stiffness is given by \cite{Lautrup2011}
\begin{equation}
    A = E_Y \frac{\pi a^4}{4}.
\end{equation}
The elastic bending forces on bead $i$ are given by 
\begin{equation}
    \bm{F}^b_i = \sum_{j=i-1}^{i+1} \pd{U_b}{\beta_j}\pd{\beta_j}{\bm{r}_i}, 
\end{equation}
so that only the nearest neighbours contribute to the bending force. We note here that although the Kratky-Porod and harmonic potentials are identical in the limit $\beta_i\to 0$, they show systematic differences for increased bending angles. As discussed thoroughly in Ref.~\cite{Bukowicki2018} for sedimenting elastic fibers and considering the Kratky-Porod and harmonic potentials, different bending potentials can lead to qualitatively different dynamics if the bending angles are large enough, and the harmonic bending potential~\eqref{eq:bendingharmonic} emerges as a preferential choice when comparing with benchmark solutions for elastic trumbbells. While the KP model is frequently used~\cite{Bernabei2013,Poier2015,Stevens2001,Frtsch2017}, other forms of the bending potential have also been studied; for example, Ref. \cite{Baroudi2019} opted for the bending potential of the form $\sum_i(\cosh\beta_i - 1)$. The selection of the relevant constitutive relation, appropriate for large deflections, is a difficult problem in general, and can only be settled with regards to a specific system.

Finally, to prevent overlap between different beds and thus self-intersections of the chain, steric interactions are typically added, such as a truncated Lennard-Jones potential between non-consecutive beads $i$ and $j$, given by
\begin{equation}
    U^{LJ}_{ij} = \epsilon \left[\left(\frac{\sigma_*}{r_{ij}}\right)^{12} - \left(\frac{\sigma_*}{r_{ij}}\right)^6\right], \quad r_{ij}<2^{1/6}\sigma_*,
\end{equation}
where the cutoff at the distance of $2^{1/6}\sigma_*$ guarantees that the potential  is purely repulsive and has no long-range attractive component.


\subsection{Continuum models}\label{sec:curve}

The elasticity of slender rods has long been a topic of interest, with many textbooks covering the most intricate aspects of the matter \cite{AudolyPomeau,Audoly2015,Powers2010}. The theory for the finite displacement of thin rods under external forces and torques was developed by Kirchhoff and Clebsch. Dill provides a detailed historical account of its development~\cite{Dill1992}. The equations of equilibrium for elastic rods, dating back to Kirchhoff \cite{Kirchhoff1859}, are non-linear even for linear elastic materials because of the complex geometry. We refer the Reader to an excellent discussion in the book of Audoly and Pomeau \cite{AudolyPomeau}, and only summarise the most important concepts here. Kirchhoff's theory encompasses the classical Euler-Bernoulli beam theory by taking up the assumption of the absence of shear deformation of the cross-sections of the deforming rod. 

A slender rod can be reduced to a curve, representing its centre line in a 3D space. However, it is more than a smooth 3D curve, since it also carries information on how much the rod twists around its centre line as we move along it. We therefore begin the description with a brief discussion of the geometric quantities involved.

\subsubsection{Geometry of a deformed rod}\label{sec:elast_fil_geom}

For simplicity, we consider here a straight rod of cylindrical cross-section and uniform mechanical properties, but this assumption can be relaxed without much difficulty. The centre line is a material curve, and we introduce the curvilinear coordinate $s$  along the centre line in a deformed configuration. To trace the elastic deformation of the rod in this configuration, we introduce the material frame $(\bm{d}_1(s),\bm{d}_2(s),\bm{d}_3(s))$ attached to the centre line and chosen such that $\bm{d}_{1,2}$ lie in the cross-sectional plane and will, in general, twist around the centre line. The vector $\bm{d}_3$ points along the curve. To describe deformation, we specify how this frame evolves along the centre line. It turns out that the derivatives of material frame vectors $\bm{d}_i$ with respect to the arc length $s$ can be expressed with the use of three scalar functions $\kappa^{(1)}(s)$, $\kappa^{(2)}(s)$, and $\tau(s)$ as
\begin{align}
    \td{}{s}{\bm{d}_1(s)} &= \tau(s) \bm{d}_2(s) - \kappa^{(2)}(s)\bm{d}_3(s), \\
     \td{}{s}{\bm{d}_2(s)} &= -\tau(s) \bm{d}_1(s) + \kappa^{(1)}(s)\bm{d}_3(s), \\
     \td{}{s}{\bm{d}_3(s)} &= \kappa^{(2)}(s) \bm{d}_1(s) - \kappa^{(1)}(s)\bm{d}_2(s).
\end{align}
These equations can be simplified upon the introduction of the Darboux vector $\bm{\Omega}(s)$:
\begin{equation}\label{eq:darboux}
    \bm{\Omega}(s) = \kappa^{(1)}(s)\bm{d}_1(s) + \kappa^{(2)}(s)\bm{d}_2(s)  + \tau(s) \bm{d}_3(s),
\end{equation}
and written as cross products
\begin{equation}\label{eq:darboux2}
    \td{}{s} \bm{d}_i(s) = \bm{\Omega}(s) \times \bm{d}_i(s), \qquad i\in\{1,2,3\}.
\end{equation}
This equation offers a convenient interpretation: the Darboux vector describes the rotational velocity of the material frame when the centre line is followed with unit speed. The rate of rotation of the frame around the directions $\bm{d}_{1}$ and $\bm{d}_2$ of the cross-section is quantified by \emph{material curvatures} $\kappa^{(1)}$ and $\kappa^{(2)}$, respectively. The third scalar function, $\tau$ is called the twist rate or {\it material twist} and it corresponds to the rotation of the material frame around the rod axis $\bm{d}_3$. It is important to stress that the Darboux vector is different from one that characterises a Fr\'enet-Serret frame of a smooth 3D curve, and material curvature and twist are, in general, different from the curvature and twist of a geometric curve. 

A geometric curve, defined by its position $\bm{r}(s)$, is characterised by two geometric quantities, its curvature $\kappa(s)$ and torsion $g(s)$. One can locally define an orthogonal frame of reference, composed of the tangent vector $\bm{t}$, normal vector $\bm{n}$ and binormal vector $\bm{b}$. The tangent vector is uniquely defined as
\begin{equation}
    \bm{t}(s) = \td{\bm{r}(s)}{s}.
\end{equation}
The curvature $\kappa$ is defined as the magnitude of derivative of the tangent vector along the centre line
\begin{equation}\label{eq:curvature_def}
    \kappa = \left|\td{\bm{t}(s)}{s}\right|,
\end{equation}
 The normal vector is then given by
\begin{equation}
    \td{\bm{t}}{s} = -\kappa \bm{n},
\end{equation}
and the binormal vector completes the basis, $\bm{b}=\bm{n}\times\bm{t}$. 
Using the subscript notation for derivatives, the curvature and torsion enter the Fr\'enet-Serret equations, which express the derivatives of basis vectors as
\begin{align}
    \td{}{s}\bm{t}(s) &= -\kappa(s) \bm{n}(s), \\
    \label{eq:tau}
    \td{}{s}\bm{n}(s) &= \kappa(s) \bm{t}(s) - g(s) \bm{b}(s), \\
    \td{}{s}\bm{b}(s) &= g(s) \bm{n}(s).
\end{align}
Here, we have a set of 2 quantities to characterise the curve, instead of 3 needed for the geometry of a material rod before. However, the geometric and material quantities are not completely independent, and the former can be computed from the latter. For example, for the geometric curvature $\kappa$ we have $\kappa^2 = (\kappa^{(1)})^2 + (\kappa^{(2)})^2$. 

\subsubsection{Energy of elastic deformation}

Having introduced the relevant geometric characteristics, one can now analyse the two basic modes of deformation for an elastic rod, that is flexion (quantified by curvature) and twist (quantified by torsion). The energy of an arbitrary deformation, involving a mixture of all modes, and possible non-uniform deformation along the length of the rod, is simply a superposition of the energies of these modes, in the limit of small strains. The resulting expression for the elastic energy of the rod is \cite{AudolyPomeau,Wolgemuth2000,Powers2010}
\begin{equation}\label{eq:energy}
\mathcal{E} = \int_{0}^{L} \text{d}s \left[ 
\frac{E_Y I_1}{2} (\kappa^{(1)}(s))^2 +  \frac{E_Y I_2}{2} (\kappa^{(2)}(s))^2 + \frac{G J}{2} (\tau(s))^2
\right],
\end{equation}
where $E_Y$ is the Young's modulus of the beam material, and $G$ is its shear modulus. The cross-sectional moments of inertia $I_{1,2}$ and the torsion constant $J$ characterise the geometry of the cross-section of the beam. For the simple case of a circular cross-section of radius $a$, they are given by
\begin{equation}
    I_1 = I_2 = \frac{\pi a^4}{4}, \qquad J = \frac{\pi a^4}{2}.
\end{equation}
Since they appear in the energy equation~\eqref{eq:energy} as products with relevant elastic coefficients, it is useful to introduce the bending rigidity $A$ and torsional rigidity $C$, given by
\begin{equation}
    A_{1,2} = E_Y I_{1,2}, \qquad C = G J.
\end{equation}

Note that the fact that the energy is quadratic in the curvatures and torsion means that, upon calculating the bending moments along the rod, they will be linearly proportional to these quantities. This fact reflects the Hookean nature of the material. More precisely, the bending moment $\bm{M}(s)$, defined as the moment of contact forces transmitted across the cross-section located at $s$, can be found as
\begin{equation}\label{eq:moment}
    \bm{M}(s) = A_1 \kappa^{(1)}(s)\bm{d}_1(s) + A_2 \kappa^{(2)}(s)\bm{d}_2(s)  + C \tau(s) \bm{d}_3(s),
\end{equation}
a relationship sometimes called the constitutive relation for the rod. 

Two important extensions are possible to the presented theory. First, the bending and twisting rigidity of body might be non-uniform. Their heterogeneity might stem either from the differential elastic properties, represented by spatially-dependent Young's modulus and shear modulus, or changing geometric properties–-–the cross-sectional shape---along the rod. In both cases, it is straightforward to generalise the energy in Eq.~\eqref{eq:energy} by introducing the dependence of the coefficients $A$ and $C$ on the arc length $s$. Second, the reference configuration of the fibre might not be straight, as is the case, e.g. for human hair or bacterial flagella. If there exists a preferential, or natural, curvature and twist, they can be incorporated into the problem by replacing the original deformation characteristics by the deviation from its natural value, yielding  
\begin{equation}\label{eq:energy2}
\mathcal{E} = \frac{1}{2}\int_{0}^{L} \text{d}s \left[ 
{A_1} (\kappa^{(1)}-\kappa^{(1)}_0)^2 +  {A_2} (\kappa^{(2)}-\kappa^{(2)}_0)^2 + {C} (\tau-\tau_0)^2
\right],
\end{equation}
where the index '0' indicates preferred values, which can also be functions of the arc length $s$ \cite{AudolyPomeau}. 

\subsubsection{Kirchhoff equations of elastic equilibrium}

The elastic energy of a rod, given in Eq.~\eqref{eq:energy}, can be used to to formulate the equations of elastic equilibrium for a one-dimensional elastic rod. As discussed in detail in Ref.~\cite{AudolyPomeau}, one way to derive them is to consider the variation of energy upon infinitesimal perturbations. This variation of energy is balanced in equilibrium by the work of distributed forces and torques. 

We consider an external force distribution, with point forces ($\bm{P}(0)$ and $\bm{P}(L)$) and torques ($\bm{Q}(0)$ and $\bm{Q}(L)$) applied to the terminal ends of the fibre, and distributed linear force and torque densities, $\bm{p}(s)$ and $\bm{q}(s)$, respectively. The distributed force $\bm{p}(s)$ can represent the weight of the rod, or non-uniform electrostatic forces if it is charged, while $\bm{q}(s)$ could be the effect of a rotating flow.   Introducing the internal force in the rod, $\bm{F}(s)$, defined as the force transmitted across the cross-section of the rod at $s$ from the upstream side ($s'\geq s$) to the downstream side ($s'\leq s$), and having defined the internal moment $\bm{M}(s)$ in Eq.~\eqref{eq:moment}, we can formulate the Kirchhoff equations as
\begin{align} \label{eq:K1}
    \td{}{s}\bm{F}(s) + \bm{p}(s) &= 0, \\ \label{eq:K2}
    \td{}{s}\bm{M}(s) + \bm{d}_3(s)\times \bm{F}(s) + \bm{q}(s) &= 0,
\end{align}
To complete the model, one must add the kinematic relationship of Eq.~\eqref{eq:darboux} and the constitutive relations of Eq.~\eqref{eq:moment}, and appropriate boundary conditions, involving the terminal values of forces, $\bm{P}$, and torques $\bm{Q}$. The Kirchhoff equations describe a balance of forces and torques on a small element of the rod. To solve them, one typically first solves for the internal force $\bm{F}$ by directly integrating Eq.~\eqref{eq:K1}. Then, one solves Eq.~\eqref{eq:K2} with the relevant boundary conditions, using the constitutive relations to eliminate the bending moments and torque in favour of the material curvatures and twist, with the orientation of the centre line determined from Eq.~\eqref{eq:darboux}. The complexity of this procedure invites numerical solutions, but analytical solutions are possible in particular geometries.

\subsubsection{Elastic forces}

To explore the density of elastic forces that arise within a deformed filament, we consider a bent and twisted configuration, where the energy functional of Eq.~\eqref{eq:energy} is recast as 
\begin{equation}\label{eq:energy22}
\mathcal{E} = \int_{0}^{L} \left[\frac{A}{2} \kappa^2  + \frac{C}{2}\tau^2 \right]\de s,
\end{equation}
where $A$ is the bending rigidity, $\kappa$ is the geometric curvature defined in Eq.~\eqref{eq:curvature_def}, $C$ is the twisting modulus of the filament, and $\tau$ is the twist density. Here and in the following, we use a shorthand notation for the derivatives with respect to the arc length denoting them by a subscript, e.g. $\bm{r}_{s} \equiv \de\bm{r}/\de s$. Note here in particular that $\kappa^2 = \bm{r}_{ss}\cdot\bm{r}_{ss}$. To ensure inextensibility of the filament, one has to take extra care when computing the variations of energy. One way to tackle this problem is to introduce tension  $\sigma(s)$ along the centre line, which may be viewed as a Lagrange multiplier guaranteeing the filament inextensibility, so that an extra term 
\begin{equation}\label{eq:incompressible}
    \mathcal{E}_\sigma = - \int_0^L \sigma(s) (\bm{r}_s\cdot\bm{r}_s) \de s,
\end{equation}
 is added to the energy~\cite{Goldstein1995,Powers2010}. Simplifications of this model not including twist can be found in Refs.~\cite{Tornberg2004,Decanio2017,CosentinoLagomarsino2005}.  
Using the principle of virtual work, the elastic force per unit length can be computed by taking functional derivatives of $\mathcal{E}+\mathcal{E}_\sigma$ as
\begin{equation}\label{eq:f_e}
\bm{f}_e = -A\bm{r}_{ssss} + C[\tau(s) (\bm{r}_s \times \bm{r}_{ss})]_s + (\sigma(s)\bm{r}_s)_s ,
\end{equation}
which is the classical Euler-Bernoulli theory. The tension needs to be adjusted so that the filament remains inextensible, as discussed in detail in~\cite{Tornberg2004}. The associated elastic moment density is found as
\begin{equation}\label{eq:M_e}
    \bm{M}_e = A \left[\bm{d}_3 \times (\bm{d}_3)_s \right] + C \tau(s) \bm{d}_3.
\end{equation}
The same procedure can be applied to an arbitrary energy density across the rod \cite{Starostin2007,Starostin2009}. We shall use knowledge about the form of elastic force density and torque density to construct various elastohydrodynamic models in the following.

We note here that, amongst other tools, there exists a discrete framework for the computation of elastic deformations of Kirchhoff rods, named the Discrete Elastic Rod (DER) model \cite{bergou2008discrete}, based on the Kirchhoff rod theory and described in detail in the book by \citet{Jawed2018}, that can be used for computations in elastohydrodynamic models. 

\subsection{Euler's Elastica}\label{sec:elastica}

A case of particular interest arises when the dynamics of an inextensible elastic rod are planar, with no twist involved. The shape of the centre line that minimises the bending energy
\begin{equation}\label{eq:eulerbendingenergy}
    \mathcal{E} = \frac{A}{2} \int_0^L  \kappa^2(s) \de s, 
\end{equation}
is referred to as Euler's elastica, for a given length and boundary conditions. The problem of finding two-dimensional shapes that minimise this elastic energy dates back to the considerations of Galileo and Hooke, and was formulated for the first time mathematically by Jacob Bernoulli. This class of problems is, however, named after Leonhard Euler, who developed the theory to study large deformations and buckling; the long history of elastica is elegantly summarised by~\citet{levien2008elastica}. 

For now, let us consider the 2D dynamics of an elastic beam that is clamped at one end, $s=0$. Instead of formulating the problem in terms of the position of the filament, $\bm{r}(s)=x(s)\bm{e}_x + z(s)\bm{e}_z$, one can equivalently use tangent angle of the filament $\theta(s)$, so that we have
\begin{equation}
    x(s) = \int_0^s\de s' \cos\theta(s'), \qquad z(s) = \int_0^s\de s' \sin\theta(s'),
\end{equation}
where we incorporated the clamping condition of $\bm{r}(0)=0$. By convention, $\theta=0$ represents a horizontal tangent, and $\theta=\pi/2$ a vertical one pointin upwards. The clamping angle is thus fixed, so $\theta(0)=\pi/2$.  For planar curves, the curvature is expressed as $\de\theta/\de s \equiv \theta_s$, so the bending energy is expressed as
\begin{equation}
      \mathcal{E} = \frac{A}{2} \int_0^L  \left(\theta_s\right)^2 \de s.
\end{equation}
In such a case, the equilibrium configuration compatible with boundary conditions at the clamped end is simply straight up, $\theta(s)=\pi/2$. Of course in dynamic problems the tangent angle becomes also time dependent, $\theta(s,t)$, but here we only mention static equilibria. One may now consider complications of the problem, such as adding a distributed weight along the rod or attaching a mass at the end, so that an additional potential energy enters the balance. In the latter case, if a mass $m$ is attached at $s=L$, it simply adds a force $\bm{P}=-m g \bm{e}_z$ acting vertically downwards. The equilibrium shape then results from the minimisation of the combined bending and potential energy of the system,
\begin{equation}
      \mathcal{E} =  \int_0^L  \left[\frac{A}{2}\left(\theta_s \right)^2 + m g \sin \theta(s) \right]\de s.
\end{equation}
The minimising shape can be found by calculus of variations, and turns out to be the solution to the ordinary differential equation
\begin{equation}\label{eq:elasticaminimal}
    {A}\theta_{ss} - mg \cos \theta(s) = 0,
\end{equation}
with two boundary conditions, $\theta(0)=\pi/2$ and $A \theta_s(L) = 0$, and which can be solved in terms of elliptic integrals. If one makes Eq.~\eqref{eq:elasticaminimal} dimensionless, the second derivative is multiplied by a dimensionless elasto-gravitational number, $A/mg L^2$, which describes the relative importance of bending stiffness vs. gravitational force, and which is the control parameter in the problem. For a detailed consideration, we refer to~\citet{AudolyPomeau}, who discuss handling additional constraints and different boundary conditions (e.g. transverse load or both ends clamped). The shapes of elastica under various conditions is a rich area of scientific work, and we refrain from discussing it in detail. However, a similar approach can be used successfully in elastohydrodynamic models.

\section{Active elastohydrodynamics of thin filaments}

The construction of elastohydrodynamic models involves coupling elastic stresses within a flexible filament with viscous stresses arising from the motion of the filament in the fluid, and possibly other forces, such as gravity or electrostatic interactions. In low-Re overdamped dynamics, the overall forces (and torques) have to balance, in agreement with Eqs.~\eqref{eq:F}–\eqref{eq:T}. 
The kinematics of the filament follow from this equation, since the velocity of the filament is encoded in hydrodynamic stresses. It turns out that the key parameter controlling the motion is the ratio of elastic to viscous forces acting on the filament. In the context of biological systems, this ratio is referred to as the Sperm number, $\text{Sp}$, introduced by \citet{Lagomarsino2003}. For a filament of length $L$ moving with a typical speed $U$ with respect to the fluid, the magnitude of viscous forces is estimated as $\zeta_\perp U$, and the sperm number is defined as
\begin{equation}
    \text{Sp} = \left(\frac{\zeta_\perp U L^4}{A}\right)^{1/4}.
\end{equation}
We bear in mind that $\text{Sp}$ incorporates the effect of fluid viscosity, since $\zeta_\perp \propto \mu$.
In passive hydrodynamics, an analogue of $\text{Sp}$ is called the compliance parameter \cite{Kurzthaler2023}, or an elasto-gravitational number \cite{li2013}, akin to that defined in Sec.~\ref{sec:elastica}. In microscale systems, another relevant parameter is the persistence length, which we define as the length $L_p$ of an elastic beam, for which the bending energy becomes comparable to the energy of thermal fluctuations of the environment, $k_B T$, $T$ being the temperature and $k_B$ the Boltzmann constant. We thus have
\begin{equation}
    L_p = \frac{A}{k_B T},
\end{equation}
which renders the reduced length $L/L_p$ another dimensionless parameter that specifies the physical context for a given system. Particles short compared to their persistence length, such as small circular DNA molecules, can be considered to be relatively stiff, with little effect of fluctuations. At the other end of the spectrum, long and weakly elastic chains, such as linkers of intrinsically disorder proteins, do not pose significant elastic resistance to their surroundings, resulting in a large conformational variability that is not affected by the elastic properties \cite{waszkiewicz2024}. 

The nonlinear coupling of elasticity and Stokesian hydrodynamics renders the field of microscale fluid-structure interactions an area of active research, with numerous applications in many areas of science, with a vast body of associated literature. The dynamics of passive flexible fibres in flows has recently been summarised in an excellent review by \citet{duRoure2019}. Here, we shall highlight selected examples that demonstrate the relevance of elasticity for microscale locomotion and transport, focusing on active systems. We note that while for passive systems bead models prove to be practical and straightforward, most problems related to activity in the context of filaments have been treated within the continuum limit. 

There are a number of dynamic phenomena associated with microscale activity that involve the effects of elasticity and hydrodynamics. Amongst them, we consider: buckling, bundling, beating, synchronisation, metachrony, swimming, pumping, and fluid-structure interaction, as keywords to organise the results.

\subsection{Buckling}

Buckling is the phenomenon of bending under load, when a system cannot sustain the arising elastic stresses and seeks a new configuration to minimise its energy. Here, we discuss two types of buckling: dynamic Euler buckling upon axial loading and rotational instability.

\paragraph{Dynamic Euler buckling}
A mechanically loaded beam undergoes Euler buckling when the load exceeds a critical threshold. In elastohydrodynamic phenomena, the source of this load are elastic stresses resulting from the motion in viscous fluid. This suggests that buckling phenomena in microhydrodynamics are of a dynamic nature, and require forcing or activity to set in. An example could be the buckled shapes of filamentous fd-viruses, which deform upon translocation through a nanopore \cite{McMullen2018}. In cytoskeletal filament motility assays, dynamic buckling of slender filaments occurs when they are driven by tangential forces into obstacles or through a crowded medium, as studied by \citet{Fily2020}. Buckling may also be caused by the growth of thin fibres which acts as an additional source of tension, e.g. as seen in liquid crystal isotropic-to-smectic-A phase transitions \cite{ShelleyUeda}. 

\begin{figure}
    \centering
    \includegraphics[width=\linewidth]{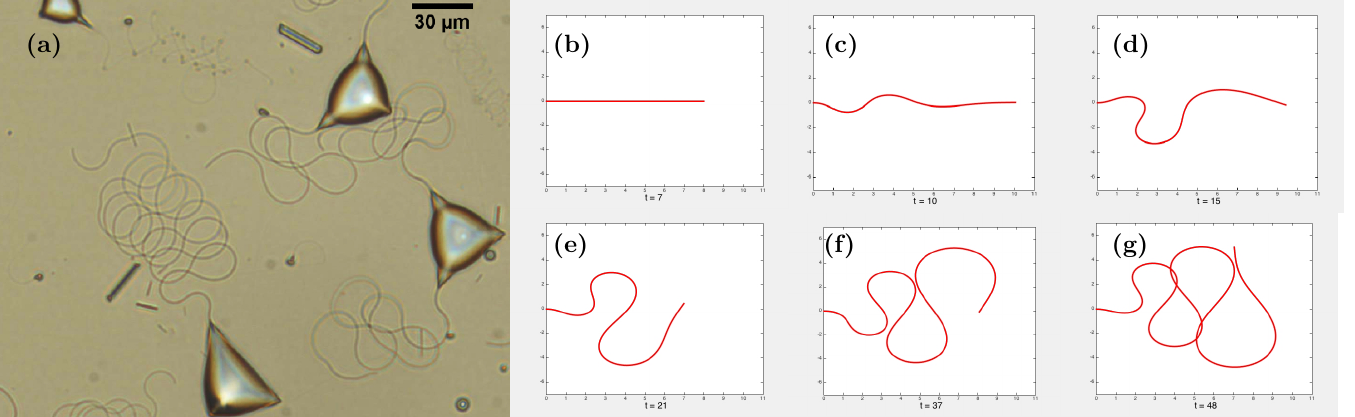}
    \caption{Buckling dynamics of elastic fibres extruded from microscale oil droplets. {\bf (a)} Upon freezing of the surfactant layer, microscopic alkane droplet extrude filamentous buckling tails. Photo credit: Stoyan Smoukov. {\bf (b)}-{\bf (g)} Snapshots from numerical solutions of the equations of motion for extruding filaments for $t\in\{7,10,15,21,37,48\}$. Initially straight, the extruded filament buckles dynamically when its length exceeds a critical threshold, and the dynamics of coiling is driven by constant extrusion of the material \cite{Cholakova2021}.}
    \label{fig:droplets}
\end{figure}

Dynamic buckling can also be driven by extrusion of an elastic filament into a viscous medium. Such a situation arises in system of self-assembled droplet microswimmers, described by \citet{Cholakova2021}. Upon cooling, microscale alkane droplets dispersed in an aqueous surfactant solution start to swim, pushed by rapidly growing thin elastic tails, seen in Fig. \ref{fig:droplets}a. When heated, the same droplets recharge by retracting their tails, swimming for up to tens of minutes in each cycle. We refrain from discussing the swimming motion here and instead focus on the dynamics of tails. They are elastic structures, filled with alkane oil, and lined with a plastic rotator layer of the freezing surfactant on the outside. Extruded at speeds of the order of 1-10 $\upmu$m/s, they create intricate spatial patterns. Their initial buckling dynamics resembles that seen in experiments by \citet{Gosselin2014}, who examined the shape of a piano string pushed into glycerine. To rationalise the buckling dynamics and longer-time patterns, we propose a simple, 2D elastohydrodynamic model. With $\bm{r}(s,t)$ describing the position of the filament, we write the energy functional as in Eq.~\eqref{eq:energy22}, with no twist contribution ($C=0$) and with an additional incompressibility condition of the filament, Eq.~\eqref{eq:incompressible}. The resulting elastic force per unit length is calculated by taking functional derivatives of the energy and reads
\begin{equation}
\bm{f}_e = -A\bm{r}_{ssss} + (\sigma\bm{r}_s)_s.
\end{equation}
We calculate viscous stresses with the aid of RFT, inverting Eq.~\eqref{eq:local_sbt}. The hydrodynamic force density acting on the filament is proportional to its local velocity, $\bm{r}_t$, written as
\begin{equation}
\bm{f}_h = -[\zeta_\parallel\bm{t}\bm{t} + \zeta_\perp\bm{n}\bm{n}]\cdot\bm{r}_t,
\end{equation}
with $\bm{t}$ and $\bm{n}$ being the tangential and normal vector of the filament, respectively. In the overdamped limit of vanishing Reynolds number, the force balance $\bm{f}_e + \bm{f}_h = \bm{0}$ yields the final equation of motion
\begin{equation} \label{fbalance}
 [\zeta_\parallel\bm{t}\bm{t} + \zeta_\perp\bm{n}\bm{n}]\cdot\bm{r}_t = -A\bm{r}_{ssss} + (\sigma\bm{r}_s)_s.
\end{equation}
A kinematic condition is now needed to reflect the fact that the filament is being pushed into the fluid at a constant speed $U$. It follows from the conservation of mass that the time derivative in the laboratory frame has the structure
\begin{equation}
\bm{r}_t \equiv \frac{\text{D}\bm{r}(t,s(t))}{\text{D}t} = \pd{\bm{r}}{t} + \pd{s}{t} \bm{t}.
\end{equation}
The first term on the right-hand side is the Eulerian velocity in the laboratory frame, while $\pd{s}{t} = U$ is the extrusion velocity. Eq.~\eqref{fbalance} can be transformed using the Fr\'enet-Serret formulae, $\bm{t}_s = -\kappa \bm{n},$ and $\bm{n}_s =  \kappa \bm{t}$, which introduce the curvature of the fibre. Upon repeated differentiation of $\bm{r}$ with respect to the arc length, we find $\bm{r}_s = \bm{t}$, $\bm{r}_{ss} = \bm{t}_s = -\kappa\bm{n}$, 
$\bm{r}_{sss} = -(\kappa\bm{n})_s = -\kappa_s \bm{n} - \kappa^2 \bm{t}$, and $\bm{r}_{ssss} =(-\kappa_{ss} + \kappa^3) \bm{n} -3\kappa\kappa_s \bm{t}$. With these relationships, we rewrite Eq.~\eqref{fbalance} as 
\begin{equation}\label{erte}
\bm{r}_t = \frac{1}{\zeta_\perp} \left[A(\kappa_{ss}-\kappa^3) - \kappa \sigma\right]\bm{n} +\frac{1}{\zeta_\parallel} \left[3A\kappa_s \kappa + \sigma_s \right]\bm{t}.
\end{equation}
The problem has no intrinsic length scale but one can be constructed out of the parameters of the fibre and its surroundings. The elastohydrodynamic length is defined as
\begin{equation}
    \ell = \left( \frac{A}{\zeta_\parallel U} \right)^{1/3}. 
\end{equation}
We make the equations of motion dimensionless using $U$ as the velocity scale, $\ell$ and the length scale, and scale the tension $\sigma$ by $A/\ell^2$. We arrive at 
\begin{equation}\label{r_dimless}
\bm{r}_t = \pd{\bm{r}}{t} + \bm{t}  = \eta^{-1} \left( \kappa_{ss}-\kappa^3 - \kappa \sigma\right)\bm{n} + \left(3\kappa\kappa_s + \sigma_s \right)\bm{t}
,\end{equation}
with $\eta = \zeta_\perp/\zeta_\parallel$. The unknowns in the problem are the shape $\bm{r}(s,t)$ and the tension $\sigma(s,t)$. We find an equation for the latter by imposing local inextensibility condition $\bm{r}_s\cdot\bm{r}_s= 0$, from which it follows that
\begin{equation}
    \pd{}{t} (\bm{r}_s\cdot\bm{r}_s) = \bm{r}_s\cdot\bm{r}_{ts}=0
\end{equation}
 Differentiating Eq. \eqref{r_dimless} with respect to the arclength, and contracting with $\bm{r}_s$ we find
\begin{equation}\label{lagrange}
\eta \sigma_{ss} - \kappa^2 \sigma = \kappa^4 - (3\eta +1)\kappa\kappa_{ss} - 3\eta \kappa_s^2,
\end{equation}
a diffusion-reaction equation for the tension. We now rewrite equations \eqref{r_dimless} and \eqref{lagrange} in terms of the tangent angle of the filament. Recalling that $\kappa=\theta_s$, we find
\begin{equation}
\bm{r}_{st} = -(\theta_t + \theta_s)\bm{n},
\end{equation} 
and we finally arrive at a set of coupled equations that describe the dynamics of the filament 
\begin{subequations}
\begin{align}\label{eq:eq1}
\eta \theta_t &= - \theta_{ssss} + [3(1+\eta)\theta_s^2 +\sigma]\theta_{ss}  + (1+\eta) \sigma_s\theta_s - \eta\theta_s, \\
\eta \sigma_{ss}-\theta_s^2 \sigma &= \theta_s^4 - (3\eta+1)\theta_s\theta_{sss} - 3\eta \theta_{ss}^2. \label{eq:eq2}
\end{align}
\end{subequations}
To complete the formulation, we need boundary conditions. We assume the clamped end of the filament, $s=0$, to be fixed, so that
\begin{equation}
    \theta(0,t) = 0.
\end{equation}
Moreover, the dimensional velocity of the clamped end is $U\bm{t}$ (or, it has zero Eulerian velocity in the laboratory frame). Evaluating Eq.~\eqref{r_dimless} and~\eqref{lagrange} at $s=0$ leads to dimensionless equations
\begin{align}
\theta_s^3(0,t) - \theta_{sss}(0,t) + \theta_s(0,t)\sigma(0,t) &= 0, \\
\sigma_s(0,t) + 3\theta_{s}(0,t)\theta_{ss}(0,t) &= 1,
\end{align}
while the free end evolves (dimensional) time according to $L(t)=U t$, so the boundary conditions need be applied at $s=L(t)$. The torque-free end means that $\bm{r}_{ss}(L(t),t)=0$, while the force condition reads $-\bm{r}_{sss}(L(t),t) - \sigma(L(t),t) \bm{r}_s(L(t),t) = 0$. Combined, these conditions can be written as
\begin{align}
\theta_{ss}(L(t),t) &= 0, \\
\sigma(L(t),t) &= 0.
\end{align}
This system of equations can be solved numerically. The shape of the filament is presented in simulation snapshots in Fig.~\ref{fig:droplets}b-g. The elastic energy is initially stored in the form of tension, and upon buckling gets partially converted into bending energy. The radius of the bends grows in time, reducing the curvature, but the extrusion supports the formation of subsequent loops. As a result, the filament exerts a force and a torque on the extruding base, and their oscillatory character can be used to construct a model of droplet locomotion to predict its swimming speed \cite{Cholakova2021}.

\paragraph{Whirling and twirling}
Another setting in which the buckling dynamics of the elastic filaments is pronounced involves rotational forcing. A straight filament, slightly tilted with respect to its rotation axis, is prone to an instability when spun \cite{Wolgemuth2000,Wada2006,limpeskin2004whirling,Maxian2022}. To quantify the nature of this instability, \citet{Wolgemuth2000} considered an overdamped continuum model of a slender filament with bending and twisting elasticity. The energy of such a filament is represented by the functional~\eqref{eq:energy2} with an additional incompressibility condition~\eqref{eq:incompressible}. The variation of energy yields the associated elastic force density, given by Eq.~\eqref{eq:f_e}. At zero Reynolds number, elastic forces balance viscous stresses, which are found from the local RFT, upon inverting Eq.~\eqref{eq:local_sbt}. This balance for a filament described by $\bm{r}(s,t)$ can therefore be expressed as
\begin{equation}
    \left({\zeta_\perp} \bm{n}\bm{n} + {\zeta_\parallel} \bm{t}\bm{t}\right)\cdot \bm{r}_t = -A\bm{r}_{ssss} + C[\tau(s) (\bm{r}_s \times \bm{r}_{ss})]_s + (\sigma(s)\bm{r}_s)_s.
\end{equation}
Similarly, the axial elastic torque, given from Kirchhoff's equations as $C\tau_s$, balances the rotational stresses, $\zeta^r \omega$, where $\omega(s,t)$ is the local angular velocity about the filament tangent vector $\bm{t}$, and $\zeta_r\approx 4\pi\mu a^2$ is the rotational drag coefficient. The dynamics are closed by a geometric constraint \cite{Powers2010}
\begin{equation}\label{eq:compatibility}
  \pd{\tau}{t} = \omega_s + (\bm{r}_s\times\bm{r}_{ss})\cdot \left(\pd{\bm{r}}{t}\right)_s,   
\end{equation}
which shows how the twist changes due to angular rotation, stretching, and out-of-plane bending motion (writhing). Using the torque balance equation $\zeta_r\omega = C\tau_s$, we finally find
\begin{equation}
    \pd{\tau}{t} = \frac{C}{\zeta_r}\tau_{ss} + \frac{1}{\zeta_\perp} (\bm{r}_s\times\bm{r}_{ss})\cdot\bm{f}_s,
\end{equation}
where the second term may be regarded as a source or sink of twist when the filament is non-planar and, at the same time, out of elastic equilibrium. To these equations, boundary conditions must be added. We now assume one end to be clamped ($\bm{r}(0)=0$ and $\bm{r}_s(0)\cdot\bm{e}_z=1$) and the other end to be free ($\bm{r}_{ss}(L)=0=\bm{r}_{sss}(L)$ and $\tau(L)=0$). These equations can be solved numerically to find the shape of the driven filament. Further analysis of the twist-bend coupling and numerical methods reveal the existence of two regimes separated by a Hopf bifurcation: (i) diffusion-dominated axial rotation, which we call twirling, and (ii) steady-state whirling motion. 

\citet{Manghi2006} studied this instability using a bead model, representing the filament as a string of spherical subunits, connected with harmonic springs and a Kratky-Porod bending potential. They included hydrodynamic interactions in the RPY approximation and applied a torque to the clamped end of the filament. They found the existence of a critical torque, beyond which a sharp discontinuous transition renders the equilibrium shape helical, with the distant end trailing behind the bent filament. They also established that this transition leads to a substantial propulsion (axial) force being exerted at the base of the filament, suggesting that this threshold might be used as force rectification device. \citet{Coq2008} designed an experimental setup to explore the shape dynamics of a spinning rod, finding a continuous but sharp transition from a straight to helical shape. Past the twirling and whirling regimes, \citet{Bruss2019} investigated further deformations of a very soft silicone fibre, which involved overwhirling, when the filament takes an 'U' shape, and the formation of plectonemes. The supercoiled structure effectively acted as Hookean storage containers for the fibre, giving the filament effective stretchability. By adding
axial flow to tension the filament, they were also able to control the onset of whirling without adjusting the material
properties of the filament, which allows to selectively engineer plectoneme formation at specified locations and lengths.

In further works, \citet{Jawed2015} investigated the propulsion and instability of a rotating flexible helical rod, combining experiments with theoretical description which accounts for the geometrically nonlinear behaviour of an elastic rod with SBT for viscous stresses, to quantify the propulsive capacity of such arrangements. The result, juxtaposed with physiological conditions for a handful of bacteria suggests that the flexibility of the flagella might impose an upper bound on propulsive force threshold, above which buckling occurs and that the buckling might provide an additional turning mechanism. Simulations of \citet{Park2017} for helices with and without an elastic hook showed that a compliant hook may control the dynamics of buckling, particularly under the rotation reversal.

\subsection{Bundling}

The topic of bundling concerns primarily bacterial filaments, whose collective rotation leads to the formation of a bundle and shapes their run-and-tumble motion \cite{Berg2004,Lauga2016}. At the root, the process of bundling is mechanical, balancing hydrodynamic flows and elastic deformations of rotating flagella. The dynamics of bundling and unbundling have been the subject of numerous studies, both theoretical or numerical \cite{Macnab1977,Kim2004,Chamolly2020, Lee2018,limpeskin2012bundling,FLORES2005,Reigh2012,Qu2018,Kong2015,Maxian2022}, and experimental \cite{Kim2004hydrodynamic,Kim2004particle,Turner2000,Turner2010,Lim2023}. Here we focus on two modelling strategies –– one continuous and one based on a discrete bead model –– to approach the subject from the two angles within the focus of this paper.

To propose a simple model of bundling, \citet{Man2017} considered two parallel straight elastic filaments of length $L$ each, rotating about their centrelines in a viscous fluid at low $\text{Re}$. The radii of the filaments are $a$, and their separation distance is $h_0$. The rotational flow they induce tends to bend the filaments around each other. The Stokes flow conditions again require the local balance of hydrodynamic and elastic force densities acting on the filaments, $\bm{f}_e(s,t) + \bm{f}_h(s,t) = 0$, and a similar torque balance. The elastic force density is derived from the instantaneous configuration of the filaments via Eq.~\eqref{eq:f_e}, which involves twist, bend, and incompressibility of the filament. In addition, for a filament rotating about its centreline with an angular velocity $\omega$, the twist density must satisfy the compatibility equation \cite{Powers2010}, which takes the form \begin{equation}
  \pd{\tau}{t} = \omega_s + (\bm{r}_s\times\bm{r}_{ss})\cdot \left(\pd{\bm{r}}{t}\right)_s,   
\end{equation}
same as Eq.~\eqref{eq:compatibility} in the case of buckling, with subscript notation for derivatives along the fibre. In addition, local torque balance implies that $C \tau_s = \zeta_r \omega$, where $\zeta_r \approx 4\pi \mu a^2 $ is the rotational friction coefficient of the rod, and this condition can be incorporated into the equation above. Finally, the relationship between elastic and hydrodynamic forces is imposed by assuming RFT to express the hydrodynamic force density for each filament as
\begin{equation}
    \bm{f}^{(i)}_h = - \left({\zeta_\parallel} \bm{t}^{(i)}\bm{t}^{(i)}+ {\zeta_\perp} (\bm{1}-\bm{t}^{(i)}\bm{t}^{(i)}) \right)\cdot\left[\bm{r}^{(i)}_t-\bm{v}^{(j)\to(i)}\right],
\end{equation}
where $\bm{v}^{(j)\to(i)}$ is the flow velocity induced by the motion of filament $j$ at the location of filament $i$. The effect of hydrodynamic interactions, embodied in this velocity, can be expressed using hydrodynamic singularities, and decomposed into rotlets, representing rotational motion of the filament centre lines, and Stokeslets for translational motion. Given the separation of scales (filament radius much smaller than filament separation, $a\ll h_0$), the two flow components are simply represented by a distribution of force and torque density on the centre lines of the filaments, and contributions from various points are integrated over the opposite filament to yield the total contribution at a given point of the filament considered. The asymptotic calculation, involving the division of integrals into local (close) and non-local (distant) contributions, yields a partial differential equation that involves two dimensionless numbers. The first is a ratio of two small parametres of the system, $\epsilon_h = h_0/L \ll 1$, and $\epsilon_a = a/L \ll 1$, essential for the asymptotic expansion, and reads $\mathcal{L}=\ln \epsilon_h/\ln \epsilon_a$. The second is the more important bundling number
\begin{equation}
    \text{Bu} = \frac{\zeta_\perp \omega a^2 L^4}{h_0^2 A},
\end{equation}
which is related to the Sperm number but additionally incorporates $\epsilon_a$ and $\epsilon_h$. One of the main results of this model is the prediction of two instabilities in the conformation of the filaments. The first instability involves a transition from weakly bent filaments to crossing, at $\text{Bu}\approx 2.1$, and from crossing to bundling at $\text{Bu}\approx 42$, on top of a strong hysteresis. The nature of these instabilities is purely mechanical, with flow-induced bending stresses that are unable to balance fluid stresses and, therefore, the fibre transits to a new configuration.  These predictions were also experimentally verified using a macroscopic design in which stainless steel wires were rotated in highly viscous silicone oil. Experimental estimates of the first instability threshold ($\text{Bu}=2.15\pm0.2$) were in excellent agreement with the theoretical prediction, while technical difficulties precluded the comparison of the second threshold. 

A natural extension of the presented model should account for the fact that the bacterial filaments are not straight but rather have a helical geometry. To rotate continuously without jamming, the flagellar filaments of bacteria need to be locked in phase  \cite{Macnab1977}. An elastohydrodynamic mechanism that leads to their synchronisation was elucidated by \citet{Tatulea2022}, who developed a rigorous procedure to coarse-grain the equations of motion in the far field using the method of multiple scales, using their earlier results on the asymptotic form of resistance matrices of two arbitrarily-shaped rigid filaments, expressed as a series expansion in reduced inverse distance $L/h_0$, with coefficients evaluated using SBT \cite{Tatulea2021}. They found that the mean phase difference $\langle{\Delta\phi}\rangle$ between two rotating rigid helices connected to the basal body by elastic hooks satisfies the celebrated Adler equation \cite{Adler1946},
\begin{equation}
\pd{\langle{\Delta\phi}\rangle}{t} = -\frac{1}{\tau_\text{sync}}\langle{\Delta\phi}\rangle    
\end{equation}
with the synchronisation time scale dependent on a dimensionless parameter being the ratio between the rotation time scale and the elastic relaxation time scale of the flagellum –– an analogue of the Sperm number.

A different class of models focusses on the mechanics of a run-and-tumble motion that incorporates the bundling process into the swimming problem. \citet{watari2010hydrodynamics} studied this problem using a bead model to elucidate the details of the trajectories of bacteria and the flow fields they produce, depending on the polymorphic form of the flagella and in the presence of geometric confinement by a wall. Microswimmers are known to interact hydrodynamically with nearby surfaces, with a famous manifestation being the circular motion of bacteria close to walls \cite{Diluzio2005}, caused by the interaction of their flow fields with their hydrodynamic images \cite{lauga2006}. In the bead-spring model of Ref.~\cite{watari2010hydrodynamics}, the cell comprises a body and multiple flagella connected with FENE-Fraenkel spring potentials to control the distances between the beads~\cite{Hsieh2006}, bending potential calculated using the bending angles, and torsional potential dependent on the torsional angles formed by four consecutive beads within a flagellum. Activity is added by imposing equal and opposite torques on the beads adjacent to the flagellar hook. The presence of pairs of torques is necessary to meet the requirement of no net force or torque on the swimmer. Hydrodynamic interactions between beads are resolved using the RPY approximation using Eqs. \eqref{eq:hydromotion} and \eqref{eq:rotne-prager-translation}, modified by an approximate treatment of the wall.  The bead-spring formulation allowed easy switching between run and tumble modes. The simulation reproduces the experimentally observed behaviour of {\it E. coli}, including run-and-tumble trajectories and clockwise circular swimming near the wall. Moreover, the polymorphic transformation of a flagellum upon reversal of rotation during a tumble was shown to facilitate the reorientation.

Lastly, we remark that bundling or twisting may emerge in macromolecular systems that are driven by their biological activity. For example, DNA twists during transcription, and the accumulation of twist density leads to writhing and the formation of supercoiled plectonemes \cite{liu1987supercoiling}, which are seen in Cryo-EM experiments \cite{Irobalieva2015}, and which influence the physical properties of DNA structure, but primarily determine its biological affinity to enzymes \cite{Vayssieres2024}. For sufficiently short DNA strands, whose length is comparable to the DNA persistence length of ca. 40-45 nm \cite{Bednar1995}, elastic models provide an insight into their structure \cite{Balaeff1999,Balaeff2006,Coleman2000}, and can be used for the prediction of its mechanical stability and hydrodynamic properties \cite{Waszkiewicz2023}. For longer DNA strands, the structure becomes more flexible, and thermal fluctuations affect the stability of writhed constructs \cite{Pyne2021}. Writhing can also be the cause of motion, when a sudden change in topology leads to elastohydrodynamic relaxation \cite{lim2008dynamics}. In addition, the topic of sequence-dependent elasticity and supercoiling of the DNA has received considerable attention in the biophysics community~\cite{Lanka2000,Sheinin2011}, including the applications of theory of elastic rods to describe sequence-dependent elasticity~\cite{Coleman2003}, it seems that there has been little progress in incorporating the effects of hydrodynamic interactions in dynamical models of this fascinating biopolymer.

\subsection{Beating}\label{sec:beating}

Eukaryotic flagella and cilia exhibit a particularly rich spectrum of beating patterns, which warrants further work to understand both single-cilia motion and collective effects that span many time and length scales~\cite{Cicuta2020,Gilpin2020}. Thus varied approaches have been proposed, from which we present a narrow selection of examples.

One of the simplest examples of elastohydrodynamic models aiming at understanding the dynamics of actuation of flexible fibres was proposed by~\citet{wiggins1998flexive} and~\citet{Wiggins1998trapping}, who considered small deformations of a flexible fibre driven by terminal forcing. In such models, if one assumes the fibre to be initially horizontal, the deflection $h(s,t)$ of the fibre obeys the following linearised dynamic equation \cite{wiggins1998flexive,Wiggins1998trapping,Camalet2000,Lagomarsino2003}
\begin{equation}\label{eq:hyperdiffusion}
    \zeta_\perp \pd{h}{t}  = - A \pd{^4 h}{s^4},
\end{equation}
earlier stated also by \citet{Machin1958}. This 'hyperdiffusion' equation has to be solved subject
to appropriate boundary conditions (corresponding to external driving). Wiggins and Goldstein discuss two fundamental modes, corresponding to Stokes' two famous problems in fluid mechanics, that is, (i) impulsive motion of an end, and (ii) terminal oscillations. For a given actuation frequency $\omega$, an elastohydrodynamic penetration length $L_\omega$ can be derived from Eq.~\eqref{eq:hyperdiffusion} as
\begin{equation}
    L_\omega = \left(\frac{A}{\zeta_\perp \omega }\right)^{1/4},
\end{equation}
from which the sperm number follows as $\text{Sp}=L/L_\omega$.

We argued before that the complexity of ciliary beating patterns stems from the distribution of active tension along the filaments. The incorporation of such distributed driving defines the next category of continuum models discussed.  In an early attempt to phenomenologically describe the beating of a cilium, \citet{Gueron1998} adopted a 2D elastohydrodynamic model derived by Gueron \& Liron \cite{Gueron1992,Gueron1993}, which combines Lighthill's SBT \cite{lighthill1976flagellar} with Kirchhoff elasticity, and showed that the experimentally observed beating pattern can result from a simple configuration-dependent forcing.

In a more complex model, \citet{Camalet2000} studied the beating and swimming of internally driven filaments in 2D by modifying the elastic beam energy \eqref{eq:energy22} to add an internal force density. Their model consists of two incompressible filaments at a constant distance $a$, attached together at one end. The pair is assumed to be held together by molecular motors and cross-linkers who generate a force $f$ per unit length that acts in opposite directions at the filaments and induces relative sliding of the pair. The relevant energy functional thus becomes
\begin{equation}\label{eq:energy23}
\mathcal{E} = \int_{0}^{L} \text{d}s \left[ \frac{A}{2}\kappa^2(s) + f(s) \Delta(s) + \sigma(s) (\bm{r}_s\cdot\bm{r}_s)
\right],
\end{equation}
where $\Delta(s) = a \int_0^s \kappa(s')\de s'$ measures the local sliding displacement, following earlier work of \citet{Everaers1995}. The relevant equation of motion emerges from the balance of viscous and elastic forces acting on the filament. Since the elastic force density found from the variation of energy, $\bm{f}_e = \delta \mathcal{E}/\delta\bm{r}$, is balanced by the hydrodynamic drag force, the local RFT approximation for the filament velocity reads
\begin{equation}\label{eq:camalet}
\pd{\bm{r}}{t} = - \left(\frac{1}{\zeta_\perp} \bm{n}\bm{n} + \frac{1}{\zeta_\parallel} \bm{t}\bm{t}\right)\cdot \frac{ \delta \mathcal{E}}{\delta\bm{r}},
\end{equation}
which is supplemented by appropriate boundary conditions \cite{Camalet2000}. They then consider small deformations to expand Eq.~\eqref{eq:camalet} in $h(s,t)$ defined earlier, to show the emergence of wave-like propagating shapes induced by a self-organized mechanism via a dynamic instability. The resulting actuation pattern can then lead to propulsion.

In the third example discussed within this Section, \citet{Eloy2012} used full, three-dimensional Kirchhoff's equations \eqref{eq:K1}–\eqref{eq:K2} to analyse the kinematics of the most efficient cilium. They considered a cilium clamped vertically to a no-slip wall, and assumed that its kinematics are known. Since the energy needed for a slender filament to produce torsion is much higher than that related to bending, they assumed no twist throughout the cilium. In this case, the Hookean relationship for the bending moment simplifies to $\bm{M}=A\bm{\Omega}$, where $\bm{\Omega}$ is the Darboux vector defined in Eq.~\eqref{eq:darboux}. Incorporating the constitutive relationship and the free-end boundary condition at $s=L$ into the Kirchhoff equations, they arrived at the following expression for the internal torque
\begin{equation}
    \bm{q}(s) = A \bm{t}_{ss} \times \bm{t} + \bm{t} \times \int_s^L \bm{F}(u)\de u,
\end{equation}
where $\bm{t}\equiv\bm{d}_3$ is the filament tangent vector, and $\bm{F}$ is the force density exerted by the filament on the surrounding fluid. To represent the overdamped dynamics in the presence of the wall, they related the instantaneous distribution of forces, $\bm{F}(s)$, to the distribution of velocity of the centreline $\bm{v}(s)$, using SBT. Within this approach, and taking into account that the presence of a rigid no-slip surface will induce interactions of the filament with its Blake's hydrodynamic image, the equation of motion becomes
\begin{equation}\label{eq:laugaeloy}
    \bm{v}(s) = \bm{\Lambda}[\bm{F}](s) + \bm{K}[\bm{F}](s) + \bm{K}^\ast[\bm{F}](s).
\end{equation}
Here, in agreement with Eq.~\eqref{eq:generalSBT}, $\bm{\Lambda}$ represents the local RFT contribution, whereas $\bm{K}$ and $\bm{K}^\ast$ are linear integral operators that account for the cilium self-interaction (Johnson's SBT \cite{Johnson1980a}) and cilium–image contribution \cite{Blake1971}. After using Legendre polynomials to diagonalise the singular part of $\bm{K}$, they discretise and invert Eq.~\eqref{eq:laugaeloy}. For a given aspect ratio of the filament and the value of $\text{Sp}$, they then seek the kinematics that maximise the pumping efficiency, proportional to the flow rate generated by the cilium, and inversely proportional to the mean expedited power. They used the DER framework \cite{bergou2008discrete} to discretise the rod and parametrise the kinematics by imposing the curvatures at discrete nodes. They then used sequential programming algorithm (SQP) to solve the optimisation problem. In two dimensions, they found that this approach leads to kinematics with the experimentally observed two-stroke cycle. During the power stroke, the cilium is almost straight, while the recovery stroke involves large curvatures. In three dimensions, they found the optimal motion to resemble that of nodal cilia, seen in embryonic development \cite{Hirokawa_2009}. The optimal kinematics depends crucially on $\text{Sp}$, with the resulting beating patterns visually resembling those observed {\it in vivo}, however, the model becomes ill-posed for large $\text{Sp}$, corresponding to very flexible filaments. 

In a similar spirit, \citet{Lauga2013} explored the shape of a planar deforming flagellum, forced internally in a periodic way. The deformations take the form of travelling waves, with the optimum being sought as the shape that minimises the energetic cost for a given swimming speed. To estimate the cost, they considered the power transmitted by the molecular motors to the flagellum, which is then only partly dissipated in the fluid. As a result, they found a family of shapes, parametrised by the Sperm number, here defined using the wavelength $\lambda$ and the period $\tau$ of the travelling wave as $\text{Sp} = (\zeta_\perp \lambda^4 / \tau A)^{1/4}$. Interestingly, for the typical value of the Sperm number for microswimmers estimated to be $\text{Sp}\approx 2-7$ \cite{VelhoRodrigues2021}, the aspect ratio of wave amplitude $h$ to wavelength, $h/\lambda$, predicted to be optimal ($h/\lambda=0.163$ for $\text{Sp}=4$ and $h/\lambda=0.194$ for $\text{Sp}=6$) seems to be close to $h/\lambda=0.188$ found by fitting to the morphological data reported in the literature for 23 different species of flagellated eukaryotes and 28 different species of spermatozoa \cite{Lisicki_Lauga_2024}.

A distinct category of microhydrodynamic problems involves the presence of follower forces. The notion of a force as a force that "follows" the geometry (retains its orientation with respect to a filament at a particular point at all times) is known from classical analyses of mechanical stability of structures \cite{bolotin1963nonconservative}. The follower-force problem is intrinsically non-variational, calling for direct simulations of the dynamics. This idea was recently applied to continuum modelling of eukaryotic flagellar beating \cite{Bayly2016}, where it is proposed as a possible mechanisms for the emergence of the observed waveform. In the same spirit, \citet{Decanio2017} examined the 2D elastohydrodynamic equations of motion of a filament with a tangential follower force applied at the terminal end of a clamped filament and found a Hopf bifurcation that sets in at high forcing, and leads to a flapping motion akin to high-Re oscillations of a free-ended garden hose. In their approach, the dynamics of the filament are governed by the Euler-Bernoulli elastic stresses, given by Eq.~\eqref{eq:f_e} with no twist ($C=0$), coupled to the fluid motion via local RFT. The resulting equation of motion for the filament becomes, similarly to the previous cases,
    \begin{equation}\label{eq:decanio}
\left({\zeta_\perp} \bm{n}\bm{n} + {\zeta_\parallel} \bm{t}\bm{t}\right)\cdot\left[\bm{r}_t-\bm{u}\right] = - A \bm{r}_{ssss} - (\sigma \bm{r}_s)_s,
\end{equation}
where $\bm{r}_t \equiv \partial\bm{r}/\partial t$ and $\bm{u}$ is a  background flow.
Upon parametrisation by the tangent angle, the equation above yields two coupled equations: one for the temporal evolution of $\theta(s,t)$ and one for the tension $\sigma(s,t)$, which they solve numerically to find the time-dependent shape. This flutter bifurcation was later seen to lead to a non-planar spinning of the buckled filament with at a locked-curvature configuration, followed by a second bifurcation at increased forcing, which involves a transition from spinning to planar oscillations \cite{Ling2018}. In an attempt to elucidate the collective effects that produce cytoplasmic streaming, e.g. during oocyte development in fruit fly, \citet{Stein2021} considered a system involving many filaments under the action of tangential follower forces, in a confined geometry, and saw a transition from spatially disordered flow patterns with little correlation to an ordered state that produced cell-sized vortical flow. The observed swirling instability supports the hypothesis \cite{Monteith2016} that cytoplasmic streaming in developing oocyte of {\it Drosophila melanogaster} fruit fly is driven by the fluid-structure interaction that leads to the self-organisation of the microtubule cytoskeleton. The emergence of large-scale vortical flows––called intracellular twisters––was also seen in large-scale simulations involving thousands of fibres enclosed in a spherical cavity \cite{Dutta2024}. 

Inspired by the dynamics of microtubules in solution, \citet{Man2019} analysed the case when the tangential follower force is attached at an intermediate point of the filament. This reflects the fact that in gliding assays of microtubules, motor proteins clamped to a substrate can propel microtubules along the surface by attaching to them and driving their motion tangentially.  Using a classical elastohydrodynamic model based on RFT and similar in spirit to these presented above, they explored the dynamic instabilities of elastic filaments with free ends, showing four distinct regimes of filament behaviour, including a non-trivial buckled state with locked
curvature. To aid in the interpretation of this state, they employed a simplified bead model, showing that the transition from locked-curvature to oscillatory state is governed by the internal tension in the chain. The successive instabilities found in the paper recapitulate the full range of experimentally-observed
microtubule behaviour, implying that neither structural nor actuation asymmetry are needed to elicit the rich repertoire of motion.

Bead models can also be profitably used to describe the motion of individual cilia. \citet{KimNetz2006} considered a bead model of an active flagellum, involving the harmonic stretching and Kratky-Porod bending potentials to represent the structure. They performed Brownian Dynamics simulations \cite{Ermak1978}, taking into account hydrodynamic interactions between beads and the effect of the wall, as well as Brownian thermal fluctuations. As a minimal model of driving, they imposed a two-phase periodic torque at the base of the cilium: a weak torque that drives the filament until a threshold inclination angle is reached, and a strong torque that drives it back. Instead of inspecting the beating, they focused on quantifying the pumping of the fluid along the surface and showed that optimal pumping is obtained for a characteristic ratio of forward-backward driving torques and Sperm number. Moreover, they observed that for independently driven filaments, hydrodynamic coupling leads to autonomous phase locking, significantly enhancing the pumping efficiency. In fact, there are cases in which bead models can be a natural choice when artificial flagella are involved. \citet{Dreyfus2005} proposed a synthetic design of a flagellar swimmer, in which the filament is composed of coated magnetic particles, connected together with DNA strands. The filament was attached to a red blood cell and actuated by an oscillating field that induced beating of the filament, resulting in propulsion of the composite design. 

Many more examples could be shown to illustrate the plethora of dynamic phenomena involving driven filaments, yet they build on the same idea of coupling elastic deformations to hydrodynamic drag forces using a simplified description that builds on the slenderness of the filaments. With a multitude of actuation modes, the behaviour of active filaments can be quantified and predicted with high precision, rendering these techniques highly useful \cite{Manghi2006hydrodynamic}.

\subsection{Synchronisation and metachrony}

Synchronisation is an ubiquitous phenomenon in nature and a generic behaviour of complex systems \cite{Strogatz,Golestanian2011}, defined as the spontaneous uniformity of phases and/or frequencies of two or more interacting oscillators. The simplest theoretical models of synchronisation describe the evolution of the phase $\phi_i$ of the $i$-th oscillator as a result of mutual interactions, so that
\begin{equation}
    \td{\phi_i}{t} = \omega_i - \sum_{j\neq i} G(\bm{r}_i-\bm{r}_j) V'(\phi_i-\phi_j),
\end{equation}
where $\omega_i$ is the intrinsic frequency. In the interaction term, $G(\bm{r})$ is the interaction kernel that describes the spatial range of interaction, and $V(\phi)$ is a periodic "potential", which drives the system towards a fully synchronised state with all $\phi_i$ being equal. The nature of the coupling depends on the underlying physical mechanism. The hydrodynamic interaction is long-range, with $G\propto 1/r^\alpha$, and $\alpha$ depending on the geometry. Notably, hydrodynamic interactions alone are not sufficient to induce synchronisation, which becomes possible when the system exhibits compliance, e.g. by adding flexibility. This was theoretically shown in systems of flexibly attached rotating helices that stop synchronising their motion when the attachment becomes stiff~\cite{Kim2004hydrodynamic,Reichert2005}.   

Many works focus on the direct analysis of synchronisation in particular systems, trying to resolve geometry and actuation in a possibly detailed way. We have seen examples of such modelling strategies when we discussed beating patterns in Sec. \ref{sec:beating}, where we outlined various approaches to actuation, also in relation to the formation of macroscopic flows. 

Using a continuum model, \citet{Goldstein2016elastohydrodynamic} presented an asymptotic analysis of the hydrodynamic coupling between two long and close filaments and found that the form of the coupling is independent of the microscopic details of internal actuation, pointing to a generic synchronisation mechanism at play when elastic and hydrodynamic effects are included. In the context of microorganisms, there has been a long-standing debate about whether the reason for the synchronisation observed in the cilia is hydrodynamic interaction or basal coupling mediated by the cell body. Experimental evidence in somatic cells of {\it Volvox carteri} shows that hydrodynamic interactions alone suffice to see the emergence of synchrony \cite{Brumley2014} but basal coupling is often the dominant responsible mechanism \cite{Wan2016}. 

In search for a generic synchronisation mechanism, in a study using the Kirchhoff equations coupled with regularised Stokeslets, \citet{Guo_Fauci_Shelley_Kanso_2018} found that for two driven filaments in-phase and anti-phase synchronisation modes are bistable and coexist for a range of driving parametres. A continuum elastohydrodynamic SBT model with geometric feedback of the cilium, accounting for motor protein kinetics to study phase synchronisation in a pair of filaments with waveforms ranging from sperm to cilia and {\it Chlamydomonas}, corroborated these results, finding both modes. These findings are consistent with experimental observations of in-phase and anti-phase synchronisation in pairs of cilia and flagella, pointing to the intertwined roles of hydrodynamics and mechanochemical feedback in synchronisation~\cite{Chakrabarti2019,chakrabarti2019spontaneous,Chakrabarti2022}. 

Collective effects in synchronisation, including the emergence of metachronal waves in arrays of cilia, have been studied and reviewed in numerous works \cite{Gueron1997,Gueron1999,Guirao2007,Yang2008,Elgeti2013,Elgeti2015,Solovev2022,Vilfan2006}, employing various levels of approximation to represent the structure and flow-induced interaction of multiple actuators. However, we turn to a different approach, and discuss the reduced models that reproduce the ingredients necessary for synchronisation. 

In search of minimal models for hydrodynamic synchronisation, a phase oscillator model was proposed to represent a single cilium that performs a circular motion \cite{Lenz2006,niedermayer2008synchronisation}. The neighbouring cilia are influenced by the induced velocity field. Synchronisation is possible when the cilia possess certain flexibility, which can be represented by an elastic confinement to their circular trajectory. In this approach, the forces driving the ciliary motion are prescribed and not calculated from the bending deformations of the filament, as we discussed before. The "discrete cilia model" proposed by \citet{Lenz2006} involves a spherical bead that follows a circular trajectory. Its flow field can then be represented by the Oseen tensor (in an unbounded fluid) or the Blake tensor (in a wall-bounded geometry). Following Ref.~\cite{Lenz2006}, we consider two spherical beads of radius $a$, each following a circular trajectory of radius $b$ at a height $h$ above a surface. The phase $\phi_i$ of each bead specifies its position. Each bead is driven by a constant torque or equivalently a constant tangential force $F\bm{t}(\phi_i)$, with $\bm{t}$ being the tangent vector on the orbit. The flow field generated at the position of bead 1 by bead 2 is given by $\bm{v}(\bm{r}_1)=\bm{G}(\bm{r}_1-\bm{r}_2)\cdot\bm{F}_2$, so that the evolution equation for $\phi_1$ becomes
\begin{equation}
    b\td{\phi_1}{t} = F \left[\zeta^{-1}+ \bm{t}(\phi_1)\cdot\bm{G}(\bm{r}_1-\bm{r}_2)\cdot\bm{t}(\phi_2)\right],
\end{equation}
with $\zeta$ being the friction coefficient of the bead. An analogous equation can be written for $\phi_2$ by swapping the indices. However, since $\bm{G}(\bm{r})=\bm{G}(-\bm{r})$, the right-hand side is symmetric with respect to exchanging the phases and therefore the phase difference $\Delta\phi$ remains constant. The situation changes when beads are softly anchored to their circular trajectories by harmonic potentials \cite{niedermayer2008synchronisation}, when spontaneous order emerges and, when multiple ciliary oscillators are aligned, metachronal waves form. This reduced representation, also called the rotor model, has gained immense popularity, since on the one hand its simple formulation invites analytical solutions and allows one to draw more general conclusions on the conditions for metachrony \cite{Uchida2011,Meng2021}, and on the other hand it can be realised experimentally in planar arrays of colloidal particles, controlled with optical traps tuned to provide orbital flexibility \cite{Kotar2010,Kotar2013,Maestro2018,Brumley2016,Wollin2011}. For a review of these works, see \citet{Bruot2016}. In a theoretical attempt to include the effects of non-planar geometry, \citet{Nasouri2016} placed several rotors outside a sphere and used the form of the Green's function outside a solid sphere \cite{Maul1996} to show that the natural periodicity of the geometry leads to synchronisation. \citet{Mannan2020} extended this and placed multiple rotors on the outer surface of a sphere, using regularised Stokeslets \cite{wrobel2016regularized} to represent the flow, and found robust symplectic metachronal waves that produce a swirling flow akin to that seen in algal colonies of {\it Volvox}.

\subsection{Swimming, pumping, and fluid-structure interaction}

Swimming is perhaps the most important aspect of microscale fluid dynamics, encompassing all the topics mentioned above \cite{childress1981mechanics,VOGEL2020,Lauga2020,Lauga:2009}. We refrain here from reviewing this rich and fascinating topic, focussing instead again on a few selected examples to show some of the approaches to the subject. 

One of the simplest examples, conceptually and computationally, of the use of bead models for swimming is the Najafi-Golestanian 3-sphere swimmer \cite{Najafi}. They proposed a linear swimmer composed of 3 colinear spherical beads, with individual control of the bead-bead arm lengths. The presence of two parameters that determine the swimming gait permits one to circumvent the limitations of kinematic reversibility of Stokes flow, and can lead to net propulsion, inspiring further designs. This emblematic system has been realised experimentally, with 3 colloids controlled by optical tweezers \cite{Leoni2009}. Generalisations of the swimmer with elastic arms, variable sphere sizes, addition of external flow, or complex fluid environments are summarised by~\citet{Yasuda2023}. 

Bead models prove particularly useful to represent artificial swimmers, which are often based on active colloidal matter. Several examples might be self-propelled worm-like active filaments \cite{IseleHolder2015,IseleHolder2016} or passive filaments driven by active colloids \cite{Laskar2017}, where patterns of activity dictate the character of flow, and the elasticity of colloidal chains determines the shapes and behaviour of composite particles. We refer here to the literature on artificial microswimmers, extensively reviewed by \citet{Bechinger2016}. Instead of discussing their details, we turn our attention here to a few continuum examples of the coupled problem of swimming and deformation.

Building on the simple elastohydrodynamic model of wiggling elastica, \citet{Lauga2007} addressed the problem of swimming of a cell endowed with an inert elastic flagellum, the base angle of which is varied sinusoidally in two or three dimensions by an internal mechanism. The induced bending waves propagate down the elastic tail, resulting in the forward motion of the swimmer. Mathematically, the problems of finding the shape of the filament and the kinematics of swimming must be solved simultaneously. The equation of motion of the swimmer within the RFT approximation is identical to Eq.~\eqref{eq:decanio}, which we repeat here for completeness,
\begin{equation}\label{eq:floppy}
\left[ {\zeta_\parallel} \bm{t}\bm{t}+ {\zeta_\perp}(\bm{1}-\bm{t}\bm{t}) \right]\cdot\bm{v} = - A \bm{r}_{ssss} - (\sigma \bm{r}_s)_s,
\end{equation}
with $\bm{v}$ being now the instantaneous velocity of the filament. We also have the inextensibility condition for the filament, written as 
\begin{equation}\label{eq:floppy2}
    \pd{}{t}(\bm{t}\cdot\bm{t})=0 = \bm{t}\cdot\bm{v}_s.
\end{equation}
The appropriate boundary conditions at the ends of the filament $s=\{0,L\}$ are
\begin{equation}\label{eq:floppy3}
    \bm{F}_\text{ext} = \pm \left( A \bm{r}_{sss} - \sigma \bm{r}_s\right), \qquad \bm{t}\times\left[\bm{T}_\text{ext}\times\bm{t} \pm A \bm{r}_{ss}\right] = 0,
\end{equation}
with the plus sign at $s=0$ and minus sign at $s=L$. Here, $\bm{F}_\text{ext}$ and $\bm{T}_\text{ext}$ are the external force and torque applied at the ends of the filament. The next step is to impose swimming kinematics, which is done by specifying the instantaneous velocity $\bm{U}(t)$ and angular velocity $\bm{\Omega}(t)$ of the swimmer, and expressing the local velocity of the filament in terms of these quantities. Finally, one needs to specify the external force and torque acting on the filament, which is achieved by considering a free swimmer, meaning that the total force and torque at $s=0$ must be balanced by hydrodynamic forces and torques on the body of the swimmer, whereas the forces and torques in at $s=L$ must vanish, since this is the free end of the flagellum. We consider a Cartesian coordinate system moving with the swimmer and located at the base of the elastic filament such that $\bm{e}_x$ points in the mean direction of the filament. In the model, we will additionally assume small deflections of the filament, described by two displacement functions, $y(x,t)$ and $z(x,t)$ in the body frame. By symmetry, the drag force and torque on the axisymmetric cell body, evaluated at its centre of resistance \cite{KimKarrila}, are given by
\begin{align}
    \tilde{\bm{F}}_\text{ext} &= -\left[\zeta^{tt}_\parallel \bm{e}_x\bm{e}_{x} + \zeta^{tt}_\perp (\bm{1}-\bm{e}_x\bm{e}_{x}) \right]\cdot \tilde{\bm{U}}, \\
    \tilde{\bm{T}}_\text{ext} &= -\left[\zeta^{rr}_\parallel \bm{e}_x\bm{e}_{x} + \zeta^{rr}_\perp (\bm{1}-\bm{e}_x\bm{e}_{x}) \right]\cdot \tilde{\bm{\Omega}},
\end{align}
with the notation corresponding to the elements of the resistance matrix of Eq.~\eqref{eq:frictionmatrix}. We emphasize here that when transforming to a different frame of reference, the form of the friction tensor changes \cite{KimKarrila} and translation-rotation coupling might appear even for highly symmetric objects. This is the case here, when the above equations are transformed to the frame we defined as reference, and they take the form 
\begin{align}
    {\bm{F}}_\text{ext} &= -\left[\zeta^{tt}_\parallel \bm{e}_x\bm{e}_{x} + \zeta^{tt}_\perp (\bm{1}-\bm{e}_x\bm{e}_{x}) \right]\cdot {\bm{U}} + a \zeta^{tt}_\perp (\bm{e}_y\bm{e}_z - \bm{e}_z \bm{e}_y)\cdot\bm{\Omega} , \\
    {\bm{T}}_\text{ext} &= - a \zeta^{tt}_\perp (\bm{e}_y\bm{e}_z - \bm{e}_z \bm{e}_y)\cdot\bm{U} -\left[\zeta^{rr}_\parallel \bm{e}_x\bm{e}_{x} + (\zeta^{rr}_\perp+a^3 \zeta^{tt}_\perp) (\bm{1}-\bm{e}_x\bm{e}_{x}) \right]\cdot {\bm{\Omega}}, \end{align}
with $a$ being the distance from the body centre to the filament attachment point, and the $tr$ and $rt$ coupling terms stated explicitly. Using these equations to impose the force-free and torque-free condition, we find the system of 9 equations for the unknown quantities $\{\bm{U},\bm{\Omega},\sigma,y,z\}$, and a set of boundary conditions. The main outcome of the model are analytical formulae that allow one to determine the performance of the swimmer and its dependence on the body shape and actuation frequency, and optimise these quantities. We shall not dive into the details of the calculation, but the formulation of the model shows that swimming problems add an additional layer of complexity to the consideration, where the kinematics become intertwined with elastohydrodynamic equations of motion.

Several other optimisation problems pertaining to swimming have been solved using a similar strategy. \citet{Spagnolie2011} considered the 12 different polymorphic forms of helical bacterial filaments and calculated their hydrodynamic efficiency. Assuming the cell to swim along the unit vector $\bm{e}_x$, the swimming velocity is written as $\bm{U}=U\bm{e}_x$, and its angular velocity reads $\bm{\Omega}=\Omega\bm{e}_x$. The corresponding fluid force and torque on the cell body are denoted by $\bm{F}=-A_0 U \bm{e}_x$ and $\bm{T}=-D_0\Omega\bm{e}_x$. If the flagellum rotates with an angular velocity $\bm{\omega}=\omega\bm{e}_x$, then the linear mobility relation for the flagellum can be written as
\begin{equation}
    \begin{pmatrix}
        \mathcal{A} & \mathcal{B} \\
        \mathcal{B} & \mathcal{D} 
    \end{pmatrix}\begin{pmatrix}
        U \\
        \omega
    \end{pmatrix} =  -
    \begin{pmatrix}
        A_0 U \\
        D_0 \Omega
    \end{pmatrix}, 
\end{equation}
where the hydrodynamic interactions between the cell and the flagellum have been neglected. Purcell defined the hydrodynamic efficiency for a flagellum as $E = \mathcal{B}^2/\mathcal{A}\mathcal{D}$, rationalising the notion that it should maximise the amount of translation resulting from its rotational motion \cite{Purcell1997}. To determine efficiency, one needs to evaluate the resistance coefficients $\mathcal{A}$, $\mathcal{B}$, and $\mathcal{D}$, which can be done using Johnson's SBT \cite{Johnson1980a}.  Equation~\eqref{eq:generalSBT} for the translating and rotating filament is written as
\begin{equation}
\label{eq:generalSBT2}
	8\pi\mu [\bm{U} + \bm{\omega}\times(\bm{x}(s)-\bm{x}_0)]= -\bm{\Lambda}[\bm{f}](s) - \bm{K}[\bm{f}](s). 
\end{equation}
with $\bm{x}(s)$ being the position of the filament centreline and $\bm{\omega}$ measured about a point $\bm{x}_0$. It may be solved using a Galerkin method to find the force density $\bm{f}(s)$. Having the force density, one can calculate the total force and torque
\begin{equation}
    F' =\bm{e}_x\cdot \int_0^L \bm{f}(s)\de s,\qquad  T' = \bm{e}_x\cdot\int_0^L (\bm{x}(s)-\bm{x}_0)\times\bm{f}(s)\de s.
\end{equation}
Now, considering a purely translational problem with $(U,\omega)=(1,0)$, one finds $\mathcal{A}=F'$, while setting pure rotation, $(U,\omega)=(0,1)$, one recovers $\mathcal{B}=F'$ and $\mathcal{D}=T'$ and the calculation of efficiency is straightforward. Comparing with available literature data on flagellar morphology, Spagnolie and Lauga conclude that the normal form is the most hydrodynamically efficient of the 12 forms by a significant margin, a result that holds sway for both peritrichous and monotrichous (polar) flagellar arrangements.

We only briefly mention here another fascinating topic that concerns the elastohydrodynamics of swimming of spermatozoa. Mammalian sperm cells have been found amongst the first examples of reported synchronised motion \cite{rothschild1949measurement} and, for their relevance to reproductive health, have been the subject of thorough investigation, including purely theoretical work devoted to their active propulsion \cite{Gaffney2011,Gaffney2021,Smith2009,Fauci2006,Gillies2009}, the interactions with each other \cite{Tung2017,Taketoshi2020,Creppy2015} and with their complex environment \cite{Elgeti2010,Ishimoto2018,FAUCI1995}.

Pumping is inherently related to swimming, being in fact another face of the same coin. Actuation of the fluid may lead to propulsion or, when a swimmer is trapped, give rise to an enhanced flow that may be used for transport. On the level of individual organisms, programmed activity of the filaments may enhance the desired biological function~\cite{Krishnamurthy2023}, while multiple flagella may lead to the emergence of collective motion in bacterial suspensions~\cite{Chen2017}, and large-scale flows upon self-organisation in bacterial colonies~\cite{Xu2019}, in active carpets comprising adhered microswimmers~\cite{Mathijssen2018,GuzmnLastra2021} or in arrays of cilia \cite{Ding2014,Boselli2021}. In addition, transport by ciliated surfaces is of primary importance in the context of nutrient delivery in bacteria-covered surfaces~\cite{Buchmann2015,Darnton2004}, but also in higher organisms, where surface-activated flow contributes to the transport of biological fluids, e.g. in the respiratory tract during mucociliary clearance \cite{Fulford1986,MVanaki2020,gsell2020hydrodynamic,Smith2008}, or in the early stages of embryonic development \cite{Smith2019}.

\section{Summary}

The paper summarises and extends some of the notions introduced during the 2023 Geilo School. The intention was to provide an introduction to the topic of elastohydrodynamics of active systems that can serve as a starting point for considerations on theoretical models used to represent flows around and shapes of microscopic slender structures. In this regard, we merely scratched the surface. However, the selected examples from a spectrum of topics related to microscale activity of slender, fibrous structures demonstrate a unified approach to coupling elastic stresses with motion in the fluid, that can be applied to analyse new phenomena in biological fluid dynamics and soft matter systems.

\bigskip

 \textbf{Acknowledgments} The Author would like to express his gratitude to the Organisers of Geilo School 2023 for the opportunity to indulge in inspiring and exciting scientific discussions in the wonderful environment of Geilo and its snow-covered surroundings. We thank Marcos F. Velho Rodrigues for the permission to use his drawings of microswimmers. Gabriela Niechwiadowicz is gratefully acknowledged for sharing her sketches of flow singularities and streamlines. 

\bigskip

\textbf{Funding} The work was supported by National Centre of Science of Poland grant no. 2018/31/D/ST3/02408 to ML.

\providecommand{\noopsort}[1]{}

\end{document}